\newif\ifMNRAS
\MNRASfalse 

\PassOptionsToPackage{pdfpagelabels=false}{hyperref} 

\ifMNRAS
    \documentclass[fleqn,usenatbib,useAMS]{mnras}
    \usepackage{graphicx}
\else

\documentclass[]{aastex63}
\fi

\usepackage{acronym}
\usepackage{hyperref}
\usepackage{amsmath,amsfonts,amssymb}
\usepackage{mathrsfs}
\usepackage{mathtools}
\usepackage[utf8]{inputenc}

\usepackage[normalem]{ulem}

\usepackage{pifont}

\newcommand{\newadam}[1]{{#1}}

\newcommand{\adam}[1]{{#1}}

\newcommand{\neweradam}[1]{{#1}}

\shorttitle{EOS CCSNe}

\begin{document}
\title{An Exploration of the Equation of State Dependence of Core-Collapse Supernova Explosion Outcomes and Signatures} 
\correspondingauthor{Aleksandr Rusakov}
\email{ar3260@princeton.edu}
\author[0009-0004-0931-0221]{Aleksandr Rusakov}
\affiliation{Department of Astrophysical Sciences, Princeton University, NJ 08544, USA}
\author[0000-0002-3099-5024]{Adam Burrows}
\affiliation{Department of Astrophysical Sciences, Princeton University, NJ 08544, USA}
\author[0000-0002-0042-9873]{Tianshu Wang}
\affiliation{Department of Physics, University of California, Berkeley, CA, 94720-7300 USA}
\author[0000-0003-1938-9282]{David Vartanyan}
\affiliation{Department of Physics, University of Idaho, ID 83843, USA}

\begin{abstract}
We explore, using a state-of-the-art simulation code in 3D and to late enough times to witness final observables, the dependence of core-collapse supernova explosions on the nuclear equation of state. Going beyond questions of explodability, we compare final explosion energies, nucleosynthetic yields, recoil kicks, and gravitational-wave and neutrino signatures using the SFHo and DD2 nuclear equations of state (EOS) for a 9-$M_{\odot}$/solar-metallicity progenitor star. The DD2 EOS is stiffer and has a lower effective nucleon mass. The result is a more extended protoneutron star (PNS) and lower central densities. As a consequence, the mean neutrino energies, final explosion energy, and recoil kick speed are lower. Moreover, the evolution of PNS convection differs between the two EOS models in significant ways. This translates in part into interestingly altered neutrino ``light" curves and noticeably altered gravitational-wave signal strengths and frequency characteristics that may be diagnostic. The faster exploding model (SFHo) yields slightly more neutron-rich ejecta and more species with atomic weights between 60 and 90 and a weak r-process.  However, this is merely a preliminary study. The next step is a more comprehensive and multi-progenitor set of 3D supernova simulations for various EOSes to late times when the observables have asymptoted. Such a future investigation will have a direct bearing on the neutron star and black hole birth mass functions and the quest towards a fully quantitative theory of supernova observables.
\end{abstract} 

\ifMNRAS
    \begin{keywords}
    stars - supernovae - general    
    \end{keywords}   
\else 
    \keywords{
    stars - supernovae - general }
\fi

\section{Introduction}
\label{sec:int}  

In the last decade, the theory of core-collapse supernova (CCSN) explosions has reached many conceptual and numerical milestones, with a consensus emerging that the turbulence-aided, neutrino-driven mechanism is the primary agency of explosion \citep{janka2012,burrows2013,wang,burrows_correlations_2024,lentz:15,2015ApJ...799....5C,2015MNRAS.453..287M,Muller2019},
with magneto-hydrodynamic (MHD) driving likely important in a subset of energetic explosions \citep{Obergaulinger2018, Obergaulinger2021,Mosta2014,mosta2015,Shankar2025,2020ApJ...896..102K}. Moreover, not only are explosions now found naturally and without artifice in detailed three-dimensional (3D) simulations, but the theory is emerging to be predictive and to address the mapping between massive-star progenitor structure and observables \citep{burrows_correlations_2024}. 

However, there is no aspect of this multi-physics numerical challenge whose uncertainties can be considered retired. The initial model structures derived from stellar evolution calculations are still in flux \citep{swbj16,sukhbold2018,laplace2021,laplace2024}. The neutrino-matter opacities, while by and large robustly known at low densities and for single particle rates, are still ambiguous at high densities due to many-body effects \citep{PhysRevC.95.025801,roberts2012,roberts_reddy2017}. Spatial grid resolution has been shown to be important \citep{hiroki_2019}, and it has not been determined that detailed 3D simulations have numerically or conceptually converged. Furthermore, the numerical neutrino transport schemes employed, though sophisticated, are not multi-angle \citep{skinner2016,skinner2019,glas2019}. Neutrino energy redistribution 
\citep{1985ApJS...58..771B,Mezzacappa2020,burrows_thompson2004,2006NuPhA.777..356B,2020PhRvD.102b3017W} by neutrino-electron and neutrino-nucleon scattering is not universally incorporated into all sophisticated codes, though it can account for $\sim$10-15\% of the neutrino heating rate in the gain region \citep{1985ApJ...295...14B}. 

Importantly, one of the most important sources of remaining theoretical ambiguity is the nuclear equation of state (EOS) employed. Its stiffness at and above nuclear densities will determine in part the compressibility, structure, and radius of the protoneutron star (PNS). Compression affects the temperature profiles, the nuclear symmetry energies affect the evolution of the electron fraction, and both are intimately tied to the neutrino transport which is central to neutrino heating in the gain region. Over the last $\sim$35 years, sophisticated nuclear equations of state that are temperature, mass-density, and electron-fraction dependent have been created and published \citep[e.g.,][]{1991NuPhA.535..331L,Shen1998,Typel2010,Steiner2013}. These stitch together the sub-nuclear-density regime with the super-nuclear-density regime to provide pressures, internal energies, entropies, chemical potentials, and species mass fractions essential for the hydrodynamics and for the neutrino-matter opacity tables. However, these equations of state and the resulting tables depend on the imperfectly known nuclear force at and above nuclear density, whose character is allied with the central open issue of the neutron star mass/radius relation. At lower densities, these equations of state are recipes that employ a medium-corrected nuclear semi-empirical mass formulae plus Saha equilibrium calculations for a single-representative heavy nucleus, alphas, and nucleons to determine the species mass fractions. 

Constraints on the resulting equations of state are few, but important, and include constraints on the nuclear symmetry terms from neutron skin measurements \citep{2017ApJ...848..105T,zhang2023bayesian,2023Parti...6...30L}, measurements of the nuclear compressibility at nuclear densities from giant monopole resonances \citep{agrawal2003nuclear}, and measurements using the {\it NICER} X-ray telescope on the International Space Station of the radii of neutron stars using light curves of X-ray spots on spinning pulsars \citep{Nicer2019,Rutherford2024}.  The former constrains the proton fraction and baryon number density dependencies of the symmetry energy to a narrow box and eliminates numerous equations of state (including \citet{1991NuPhA.535..331L}). The $^{208}$Pb data constrain the nuclear stiffness around nuclear density and eliminates
others (e.g., \citet{Shen1998}). The {\it NICER   } dataset suggests neutron star radii for neutron stars of mass near 1.4 M$_{\odot}$ between 10.5 and 13 kilometers (km). The accurate measurement of neutron star masses above the maximum neutron star mass predicted by some EOSes eliminates them, but many such limits can be weak \citep{Fan2024}.      

Most previous explorations of the role of the nuclear EOS on supernova models focused on ``explodability" \citep[e.g.,][]{Suwa2013,Couch2013,burrows2018,PhysRevLett.124.092701,Powell2025}. Some of these earlier works found either that a high nucleon effective mass \citep{Schneider2019,PhysRevLett.124.092701} or a softer (lower nuclear compressibility) EOS \citep{Couch2013} were conducive to explosion or to earlier explosion. This is one reason the LS220 and LS180 EOSes \citep{1991NuPhA.535..331L} were often shown to be more propitious, though they have now been shown to be inconsistent with nuclear systematics and laboratory measurement \citep{2017ApJ...848..105T}. This is also why the stiff Shen EOS \citep{Shen1998} had been shown to make explosions more difficult or more delayed \citep{Suwa2013}. 

\citet{Powell2025} have recently suggested that the unifying feature of these studies is the role of the EOS in leaving a more compact protoneutron star. Higher effective nucleon masses will lower the thermal component of the pressure, as will softer nuclear matter above nuclear densities.
The result in both cases will be more compact protoneutron stars. The associated compression near the neutrinospheres will translate into higher neutrinosphere temperatures. The latter will result in harder emergent neutrino spectra and, since neutrino absorption in the gain region is an increasing function of neutrino energy, the upshot is greater absorbed power where it counts. This agency is similar to the relative effect of general relativity vis \`a vis Newtonian gravity \citep{bruenn_rel}. It is also similar to what would happen if $\nu_{\mu}$, $\bar{\nu}_{\mu}$, $\nu_{\tau}$, and $\bar{\nu}_{\tau}$ emission and loss were somehow increased, paradoxically facilitating explosion while accelerating energy loss through a nearly useless channel. Moreover, a more compact core allows the accreted matter from the envelope to reach smaller radii earlier, and this will make gravitational energy available earlier at a greater rate. 

In this paper, we add to this literature by comparing 1) the evolutions and 2) many observables and behaviors of two long-term 3D simulations to the asymptotic state of one progenitor (the 9-M$_{\odot}$, solar-metallicity model of \citet{swbj16}) for two different nuclear equations of state: SFHo \citep{Steiner2013} and DD2 \citep{Typel2010}.  These two EOSes were chosen because they are still consistent with all known empirical constraints, while being different enough to be an interesting comparison. The DD2 EOS, stiff as it is, is just marginally consistent with the {\it NICER} radius data.  \adam{Moreover, \citet{das2024} disfavor it in the context of their neutron star cooling analysis.} However, it provides a good foil to the SFHo EOS. The SFHx EOS is still alive, but is too close to the SFHo EOS to allow an interestingly different span of outcomes and observables. To our knowledge there has been no previous such comparison in 3D to late times with a focus beyond mere explodability.

In \S\ref{method}, we summarize our computational methodology using the state-of-the-art code F{\sc{ornax}}. Then, we follow in \S\ref{hydro} with a discussion of the differences in the 3D hydrodynamics and explosions when these two different equations of state are embedded into our supernova code. We also include a discussion of the differences in the thermal profiles and in the differential character of protoneutron star convection. In \S\ref{lum}, we contrast the emerging neutrino luminosities and mean energies. Section \ref{energy} addresses the differences in the asymptotic explosion energies and recoil kicks, and in \S\ref{nucleo} we explore the nucleosynthetic yields. In \S\ref{gw}, we contrast the gravitational wave signatures. We close out in \S\ref{conclusions} with a recap of what we have found, followed by a general discussion of the insights gleaned and the potential next steps in the investigation of the role of the nuclear equation of state in core-collapse supernova explosions and in neutron star and black hole birth.

\section{Method}
\label{method}

For this study of how differences in the nuclear equation of state may translate into differences in the behavior and signatures of core-collapse supernovae, we use our code F{\sc{ornax}}. \citet{skinner2019} and \citet{vartanyan2018b} provide a more comprehensive discussion of the code. F{\sc{ornax}} has been exercised extensively, with many papers over the last eight years \citep{radice2017b,burrows2018,vartanyan2018a,vartanyan2018b,vsg2018,burrows_2019,radice2019,hiroki_2019,vartanyan2019,Nagakura2020,vartanyan2020,burrows_2020,nagakura2021,2021Natur.589...29B,2022MNRAS.510.4689V,coleman,burrows_40,burrows_correlations_2024,burrows_40,Channels2025}.  It sports GPU and CPU variants and has been run on the NERSC/Perlmutter, ALCF/Polaris, ALCF/Aurora, and TACC/Frontera machines.

\begin{figure*}   
    \centering   
    \includegraphics[width=0.47\textwidth]{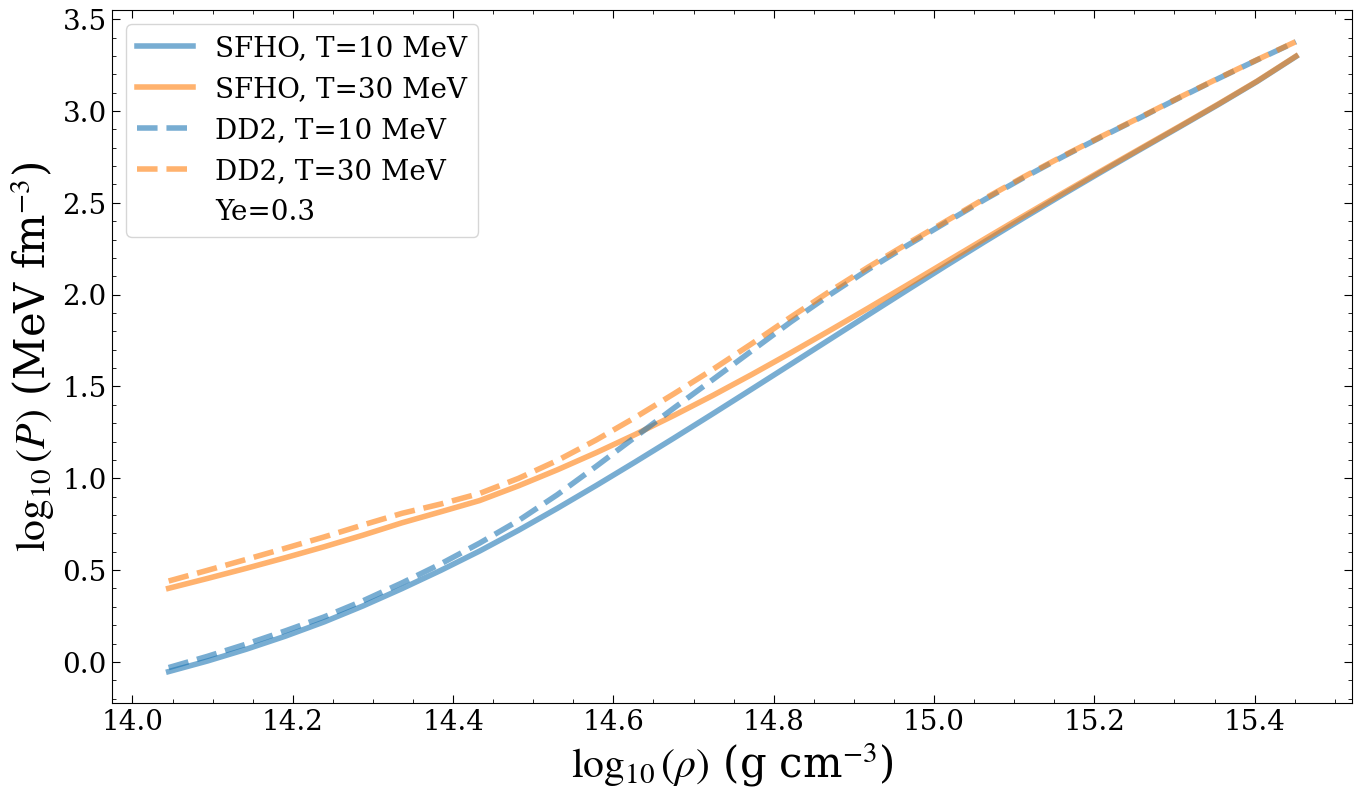}
    \includegraphics[width=0.47\textwidth]{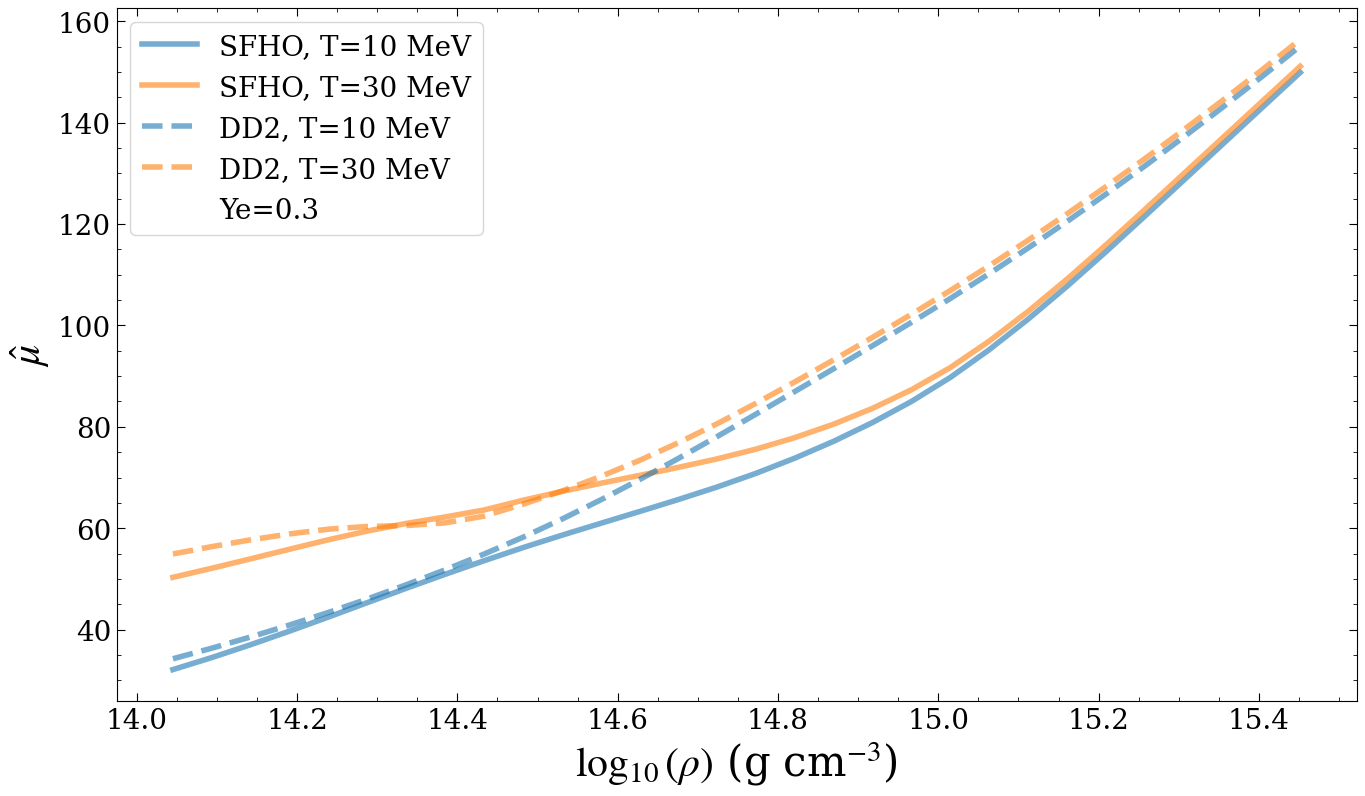}
    \includegraphics[width=0.47\textwidth]{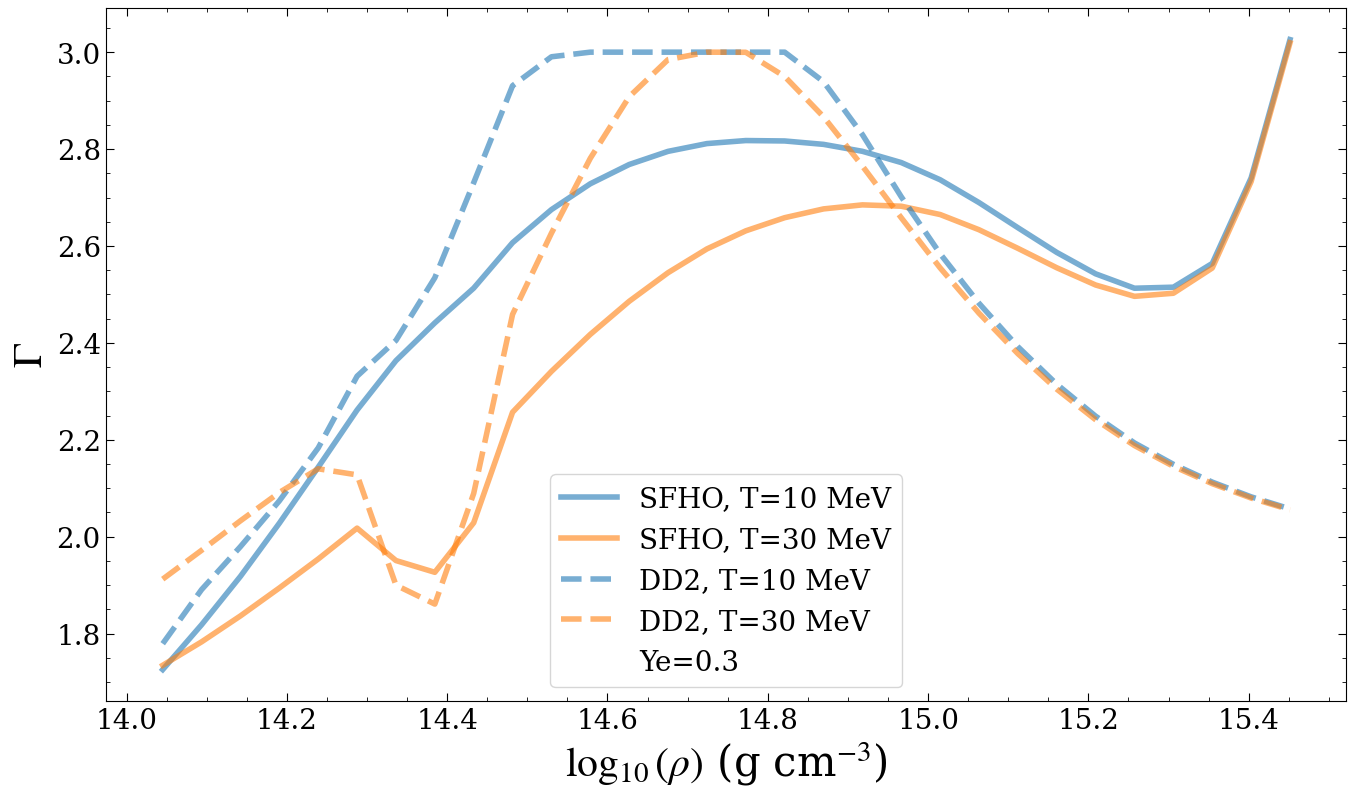}
    \includegraphics[width=0.47\textwidth]{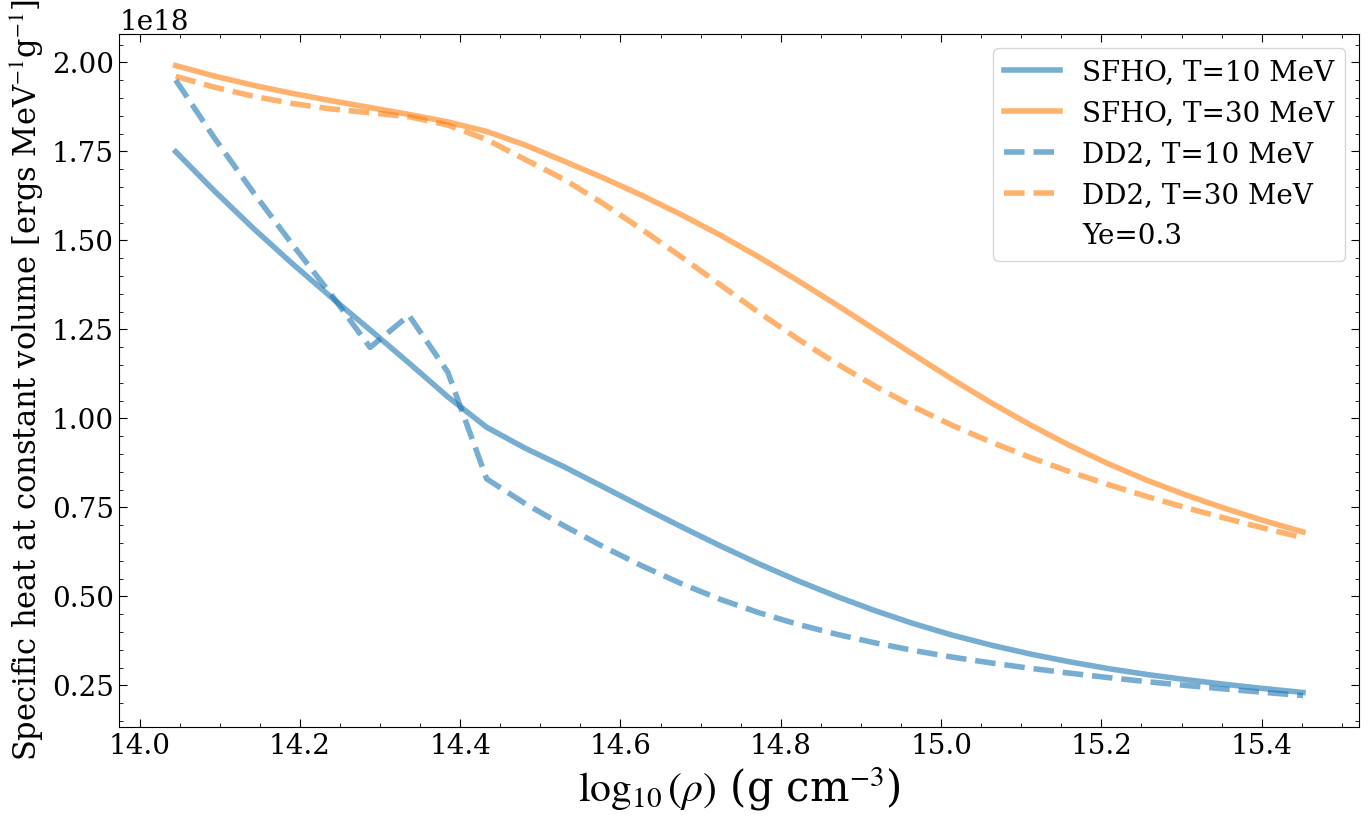}
    \caption{Comparisons of aspects of the DD2 (dashed) and SFHo (solid) equations of state. Here, we fix the $Y_e$ value at 0.3 and show pressure, $\hat{\mu}$ ($= \mu_n - \mu_p$), adiabatic gamma, and specific heat at constant volume as functions of density for two given temperatures for both equations of state.}
    \label{fig:eos_comp}      
\end{figure*}

\newadam{For these 3D F{\sc{ornax}} simulations, we use the SFHo \citep{Steiner2013} and DD2 \citep{Typel2010} nuclear equations of state. The maximum gravitational and baryonic masses of the SFHo EOS are 2.06 $M_{\odot}$ and 2.42 $M_{\odot}$, respectively, while the corresponding numbers for the DD2 EOS are 2.42 and 2.93 M$_{\odot}$. In the same order, the compressibilities ($K$) and the effective masses ($m^*$) at nuclear densities ($\sim$2.6$\times$10$^{14}$ g cm$^{-3}$) are 245.4 and 242.7 MeV and 0.76 and 0.56 (in units of the nucleon mass). However, the DD2 EOS gets stiffer faster above nuclear density. This is demonstrated in Figure \ref{fig:eos_comp} and is a central difference between the two EOSes. This is also the basic reason the cold, $\beta$-equilibrium neutron star maximum mass for the DD2 EOS is so much larger than that for the SFHo EOS\footnote{This difference will also translate into differences in the birth fraction and masses of black holes created across the continuum of massive star progenitors \citep{Channels2025}. However, this interesting observation is not the subject of the current paper. See \S\ref{conclusions}.}. In Figure \ref{fig:eos_comp}, we also compare the $\hat{\mu}$ ($= \mu_n - \mu_p$), adiabatic gammas ($\Gamma$), and specific heats for the two equations of state. The differences in the $\Gamma$s in particular recapitulate the differences in the pressures in the relevant density regimes. The differences in the specific heats in part reflect the differences in the effective nucleon mass. In addition, we note that the lower effective nucleon mass of the DD2 EOS raises the thermal component of the pressure. As a result, both the greater stiffness and lower $m^*$ result in larger-radius protoneutron stars and lower central densities. As suggested in \S\ref{sec:int}, this leads to the expectation that the use of the DD2 EOS will lead to weaker explosions.  At least for the 9.0-M$_{\odot}$ progenitor used in this study, this conclusion is substantiated.}

The 9-M$_{\odot}$ \newadam{solar-metallicity} progenitor model for this study was taken from \citet{swbj16}. \newadam{This 9-$M_{\odot}$ progenitor was chosen for this study because, due to its steep mass density profile in the Chandrasekhar core and small compactness\footnote{The compactness parameter loosely characterizes the shallowness of the mass density profile of the Chandrasekhar core of the progenitor mass star and is defined as
\begin{equation}     
\xi_M= \frac{M/M_{\odot}}{R(M)/1000\, \mathrm{km}}\,,
\end{equation}
where the subscript $M$ denotes the interior mass coordinate at which compactness is evaluated. We evaluate it at $M$ = 1.75 M$_{\odot}$.} ($= 6.57\times 10^{-3}$ at an interior mass of 1.75 $M_{\odot}$), its explosion phase is of shorter duration ($\sim$2.0 seconds) and the final observable properties (such as explosion energy, kick speed, and nucleosynthesis) asymptote more quickly than found for more massive progenitors \citep{burrows_correlations_2024}. This saves on the computational resources required. It is also the case that we have explored this progenitor in numerous other studies, enabling ready comparison \citep{burrows_2020,vartanyan2023,burrows_correlations_2024}.}

Our spatial tiling is $1024\times128\times256$ for our 3D spherical coordinate ($r\times\theta\times\phi$) grid, which extends from zero to 100,000 kilometers in the spherical radius. We use twelve neutrino energy groups for each of three species ($\nu_e$, $\bar{\nu}_e$, and ``$\nu_{\mu}$" [$\equiv$ $\nu_{\mu}$, $\bar{\nu}_{\mu}$, $\nu_{\tau}$, $\bar{\nu}_{\tau}$]), where the $\nu_e$ neutrinos are logarithmically spaced from 1 to 300 MeV and the other species are logarithmically spaced from 1 to 100 MeV. Energy redistribution by neutrino-electron scattering is handled using the moment scheme of \citet{burrows_thompson2004} and \citet{2006NuPhA.777..356B} and energy redistribution by neutrino-nucleon scattering is handled using a Kompaneets \adam{energy-diffusion} method \citep{2020PhRvD.102b3017W}. \adam{Since energy transfer to the nucleon in neutrino/nucleon scattering is a small fraction of the neutrino energy (due to the large nucleon mass), the Kompaneets method, for which one solves a diffusion-like differential equation, is a fast and efficient approach for handling energy redistribution of the neutrino spectra. Unlike for neutrino/electron scattering, for which the associated energy transfer is large and the process couples all energy groups, in the solution of the Kompaneets energy diffusion equation neutrino/nucleon scattering couples only adjacent energy groups, with the result that the numerical scheme scales only linearly in energy-group number and is quite efficient. Otherwise, as for neutrino/electron scattering, numerical schemes scale quadratically with group number and are quite expensive and relatively slow \citet{burrows_thompson2004} and \citet{2006NuPhA.777..356B}.} 

General relativity is handled approximately, using the TOV equation (with velocity and neutrino terms) \`a la the Case A scheme of \citet{marek2006}. Gravitational redshifts are included in the neutrino transport. To avoid a too stringent angular Courant hit, we deresolve in both $\theta$ and $\phi$ for small spherical radii and small cylindrical radii, respectively. This is a so-called ``dendritic" grid \citep{skinner2019}. Moreover, since the Courant constraint in the deep interior (where the speed of sound is some modest fraction of the speed of light) is already severe, we can solve the spatial operator of the transport equation
(the ``left-hand-side") explicitly, without the need for global solves. The combination of the dendritic grid and the explicit approach to the spatial operators, along with the use of the Kompaneets scheme for neutrino-nucleon redistribution, accelerates the computational speed of F{\sc{ornax}} by many factors.  The initial spins were set to zero and global angular momentum is conserved to better than 0.1\%. Coupled with generous allocations of computer time on supercomputers, this combination enables the completion to many seconds after bounce of many 3D simulations during one year and has transformed our ability to explore robustly the full landscape of CCSN possibilities.

\section{The Basic Hydrodynamic Results and Differences}
\label{hydro}

\begin{figure*}   
    \centering
    \includegraphics[width=0.65\textwidth]{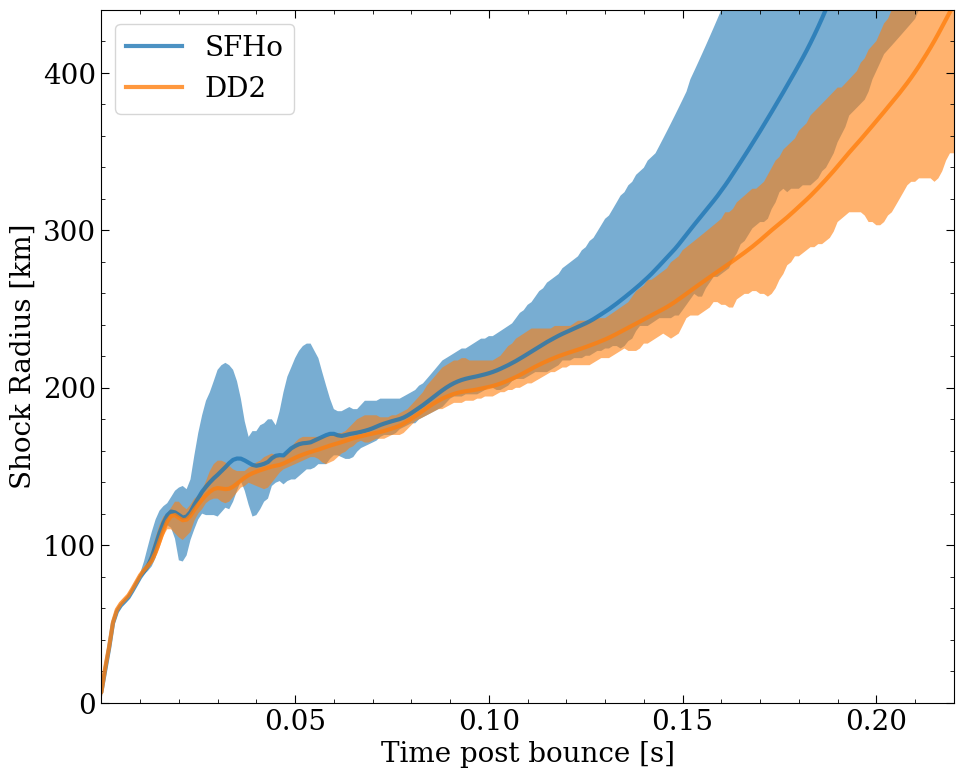}
    \includegraphics[width=0.65\textwidth]{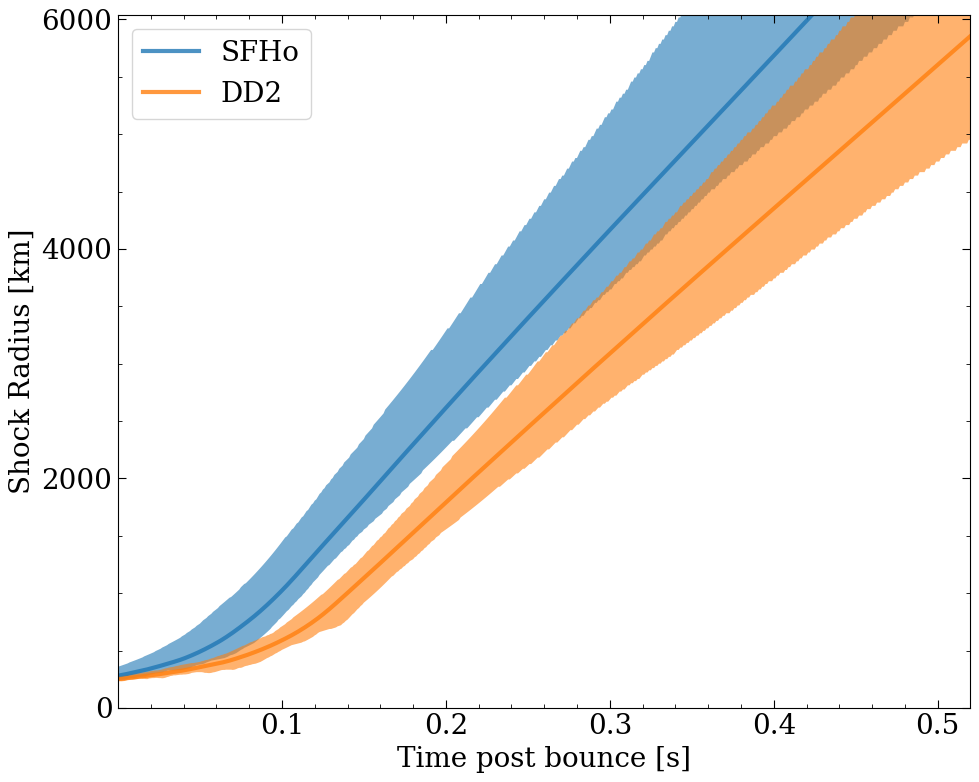}
    \caption{Mean (solid line), maximum, and minimum shock radii versus time after bounce (in seconds) for the SFHo (blue) and DD2 (orange) equations of state on a scale of 450 km (top) and a scale of 6000 km (bottom). Note the slower rise of the shock position and lower late-time shock speed of the DD2 model.}
    \label{fig:shock_r_comp}      
\end{figure*}

\begin{figure*} 
    \centering
    \includegraphics[width=0.47\textwidth]{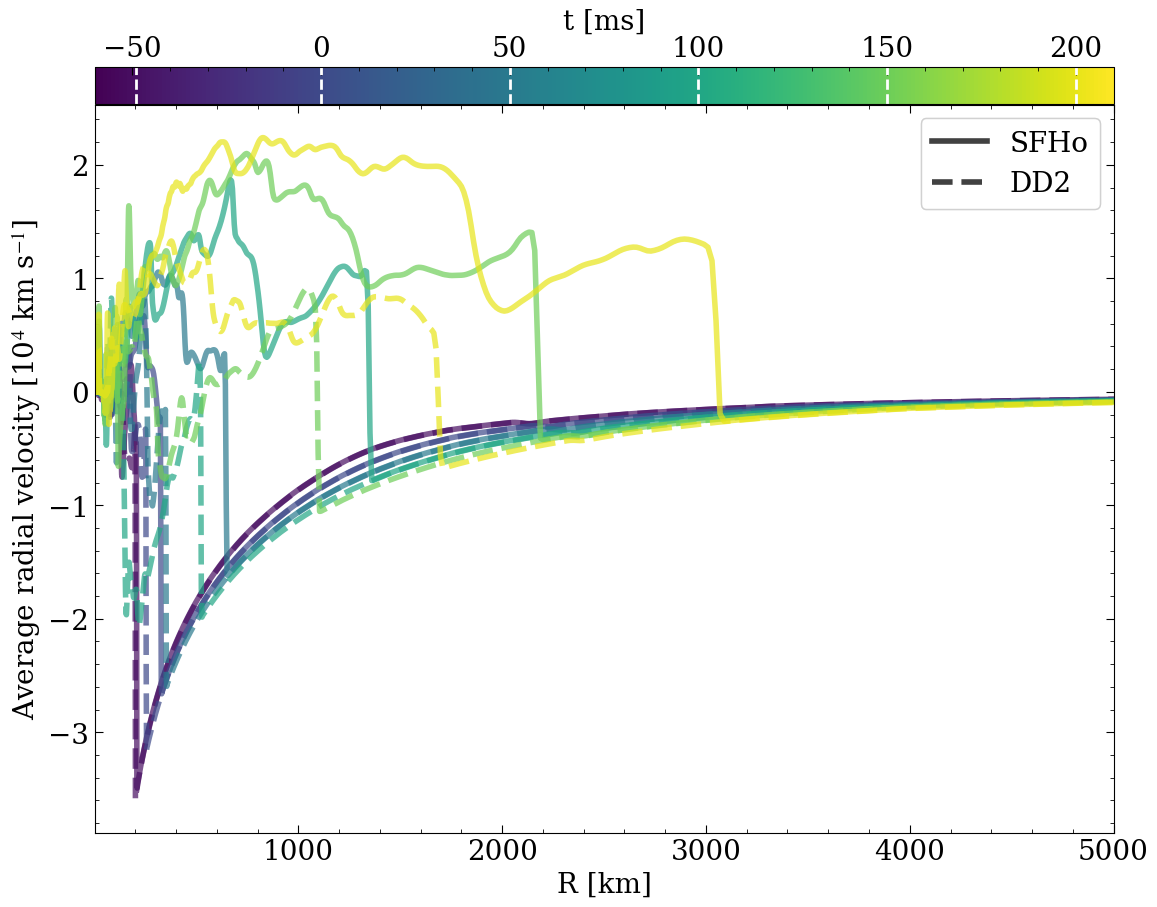}
    \includegraphics[width=0.47\textwidth]{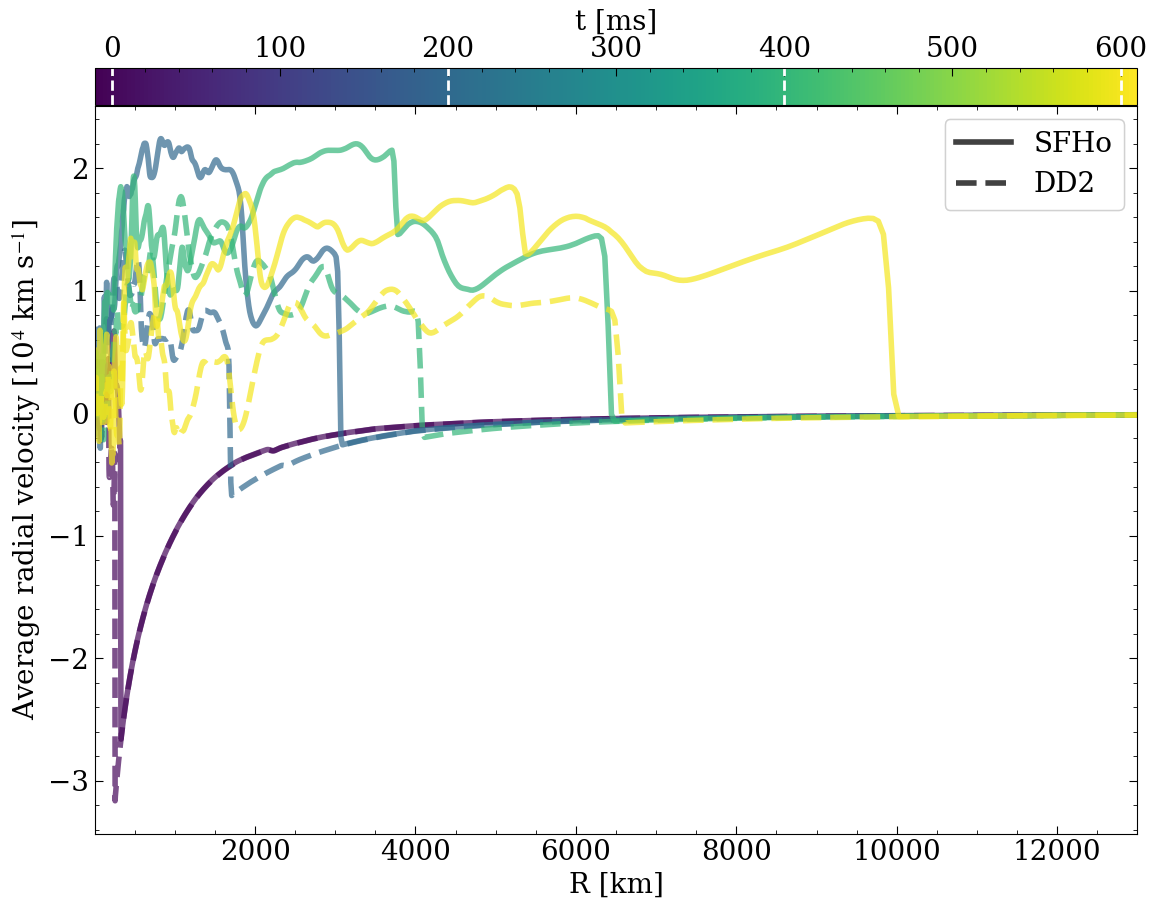}
    \caption{SFHo (solid) and DD2 (dashed) velocities versus radius (along one random radial ray \adam{in the equatorial plane of the star}) up to $\sim$200 ms after bounce (left) and $\sim$600 ms after bounce (right). \adam{The sharpness of the shock position is preserved by not averaging over the sphere.} Note that the revival of the shock is earlier for the SFHo EOS and that for it the shock and post-shock speeds are greater as well.}
    \label{fig:vel}      
\end{figure*}

\begin{figure*}    
    \centering
    \includegraphics[width=0.85\textwidth]{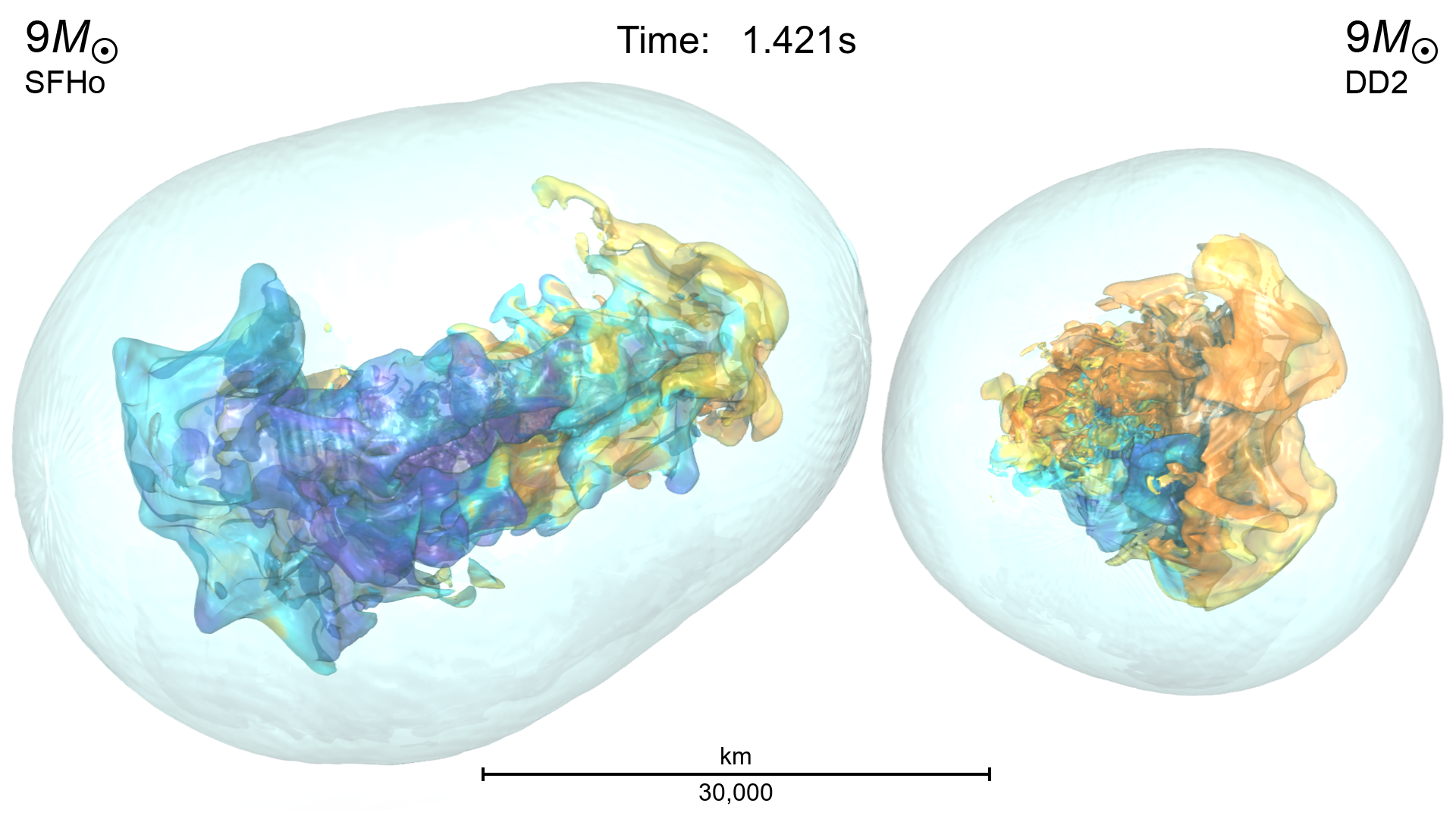}
    \caption{\adam{A comparison of snapshots of the 3D structure of the SFHo and DD2 models $\sim$1.42 seconds after bounce.  The interior colors are an isoentropy profile, colored by $Y_e$, with blue lower and yellow higher. This figure gives a more comprehensive sense of the relative shock speeds for the two models as a function of direction and demonstrates the more quadrupolar character of this SFHo explosion.  Note that the chaos of such flows mitigates against any conclusion that such a morphology might be universal.  It likely is not. However, the relative speeds of these models (SFHO $>$ DD2) seen here and in Figure \ref{fig:vel} likely qualitatively are.}}
    \label{fig:3Dshock}      
\end{figure*}

Figure \ref{fig:shock_r_comp} depicts the early evolution of 
the shock radii (mean, maximum, and minimum) for the two equations of state. As frequently witnessed \citep{burrows_2020,burrows_correlations_2024,glas2019,stockinger2020}, the 9-$M_{\odot}$ model of \citet{swbj16} explodes rather easily, barely stalling. However, as both the top and bottom panels on Figure \ref{fig:shock_r_comp} indicate, the SFHo model explodes slightly earlier and achieves a higher shock velocity.
\adam{Specifically, the shock speed is direction-dependent, and varies after explosion from $\sim$15,000 to 18,000 km s$^{-1}$ for the SFHo model and from $\sim$12,000 to 15,000 km s$^{-1}$ for the DD2 model. Note that the band of color for each model on Figure \ref{fig:shock_r_comp} reflects the maximum and minimum in shock position with solid angle.} Figure \ref{fig:vel} portrays the velocity profile in radius (along, \adam{arbitrarily, the x-axis}) at various times both before and after bounce for the SFHo(solid) and DD2 (dashed) models, reiterating for the DD2 model the later onset of explosion and the slightly lower shock and post-shock speeds.  \adam{Figure \ref{fig:3Dshock} provides a 3D rendering of the debris and shock wave (light blue veil) at $\sim$1.42 seconds after bounce, recapitulaing the stated differences between the two models in speed.} As anticipated (\S\ref{sec:int}), Figures \ref{fig:shock_r_comp}, \ref{fig:vel}, and \ref{fig:3Dshock} capture the weaker nature of the explosion for the DD2 model. This weakness tracks the lower heating rates versus time in the gain region plotted in Figure \ref{fig:Qdot_comp}.

\begin{figure*}   
    \centering
    \includegraphics[width=0.47\textwidth]{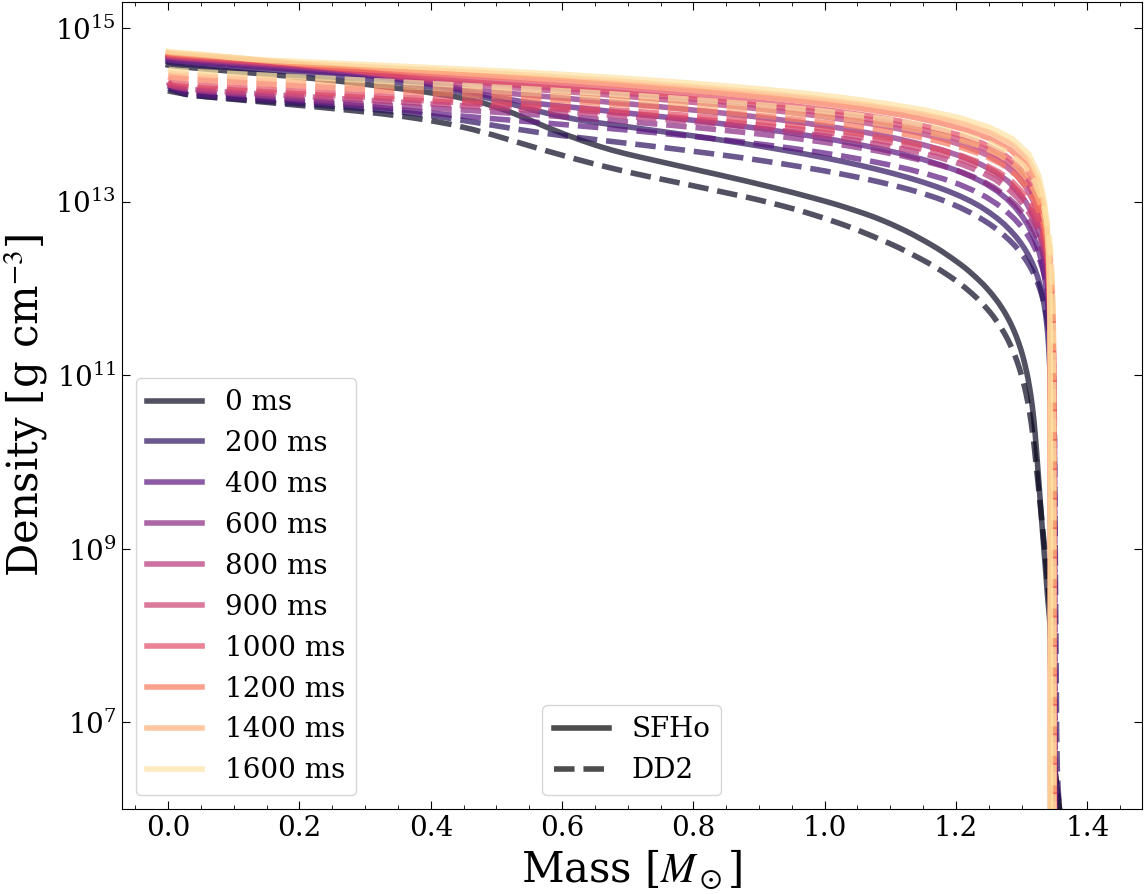}
    \includegraphics[width=0.47\textwidth]{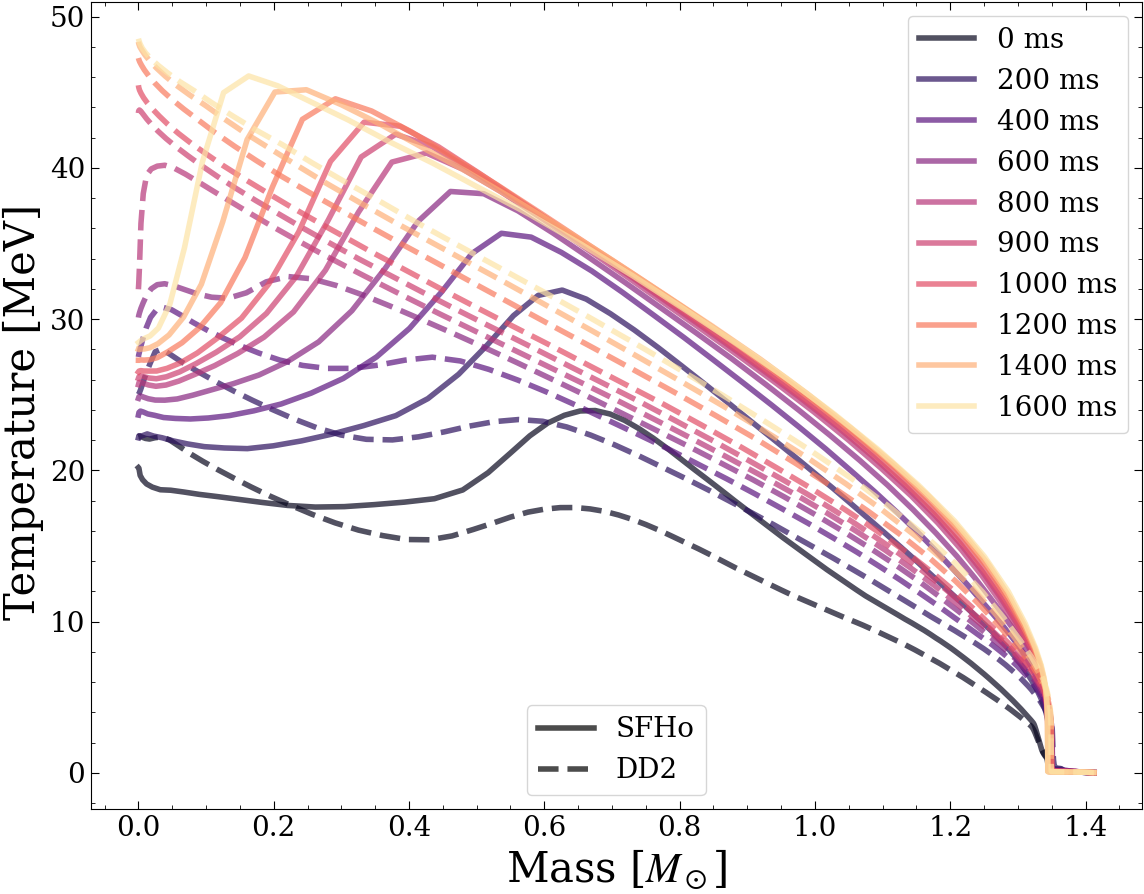}
    \includegraphics[width=0.47\textwidth]{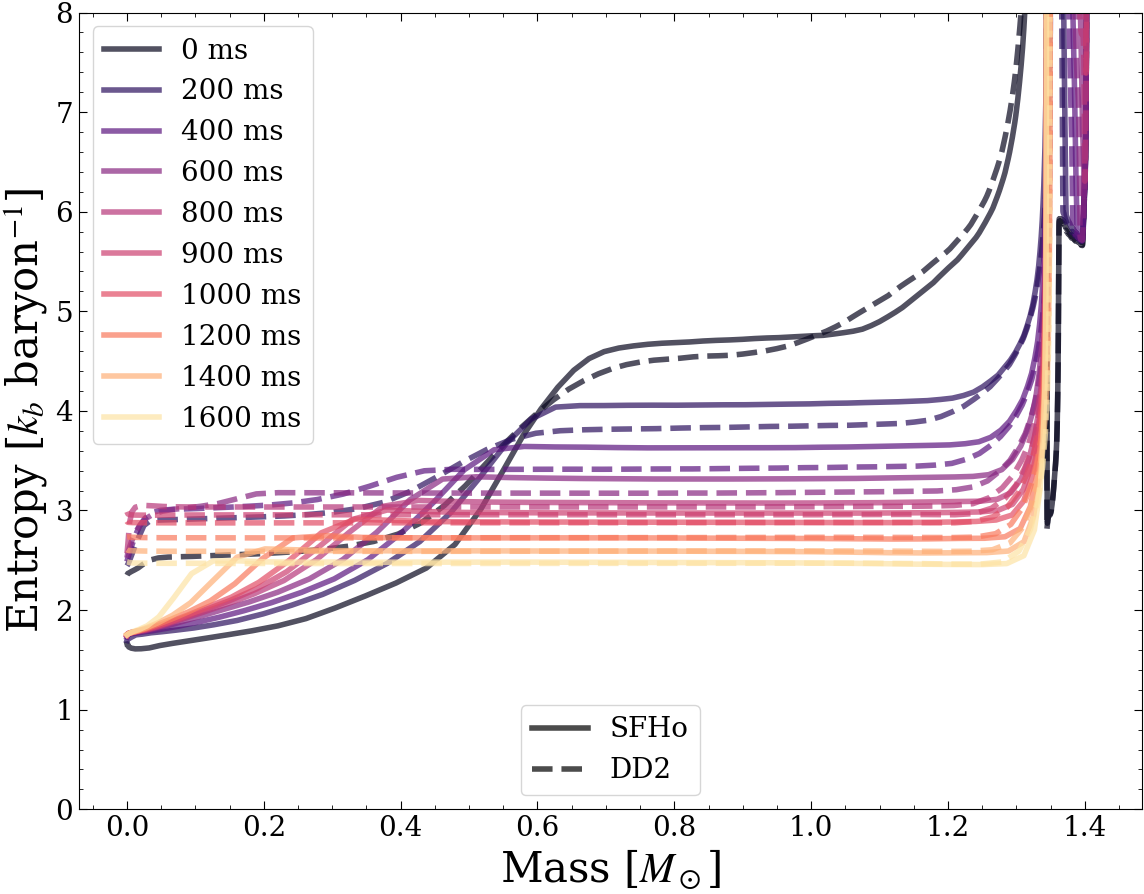}
    \includegraphics[width=0.47\textwidth]{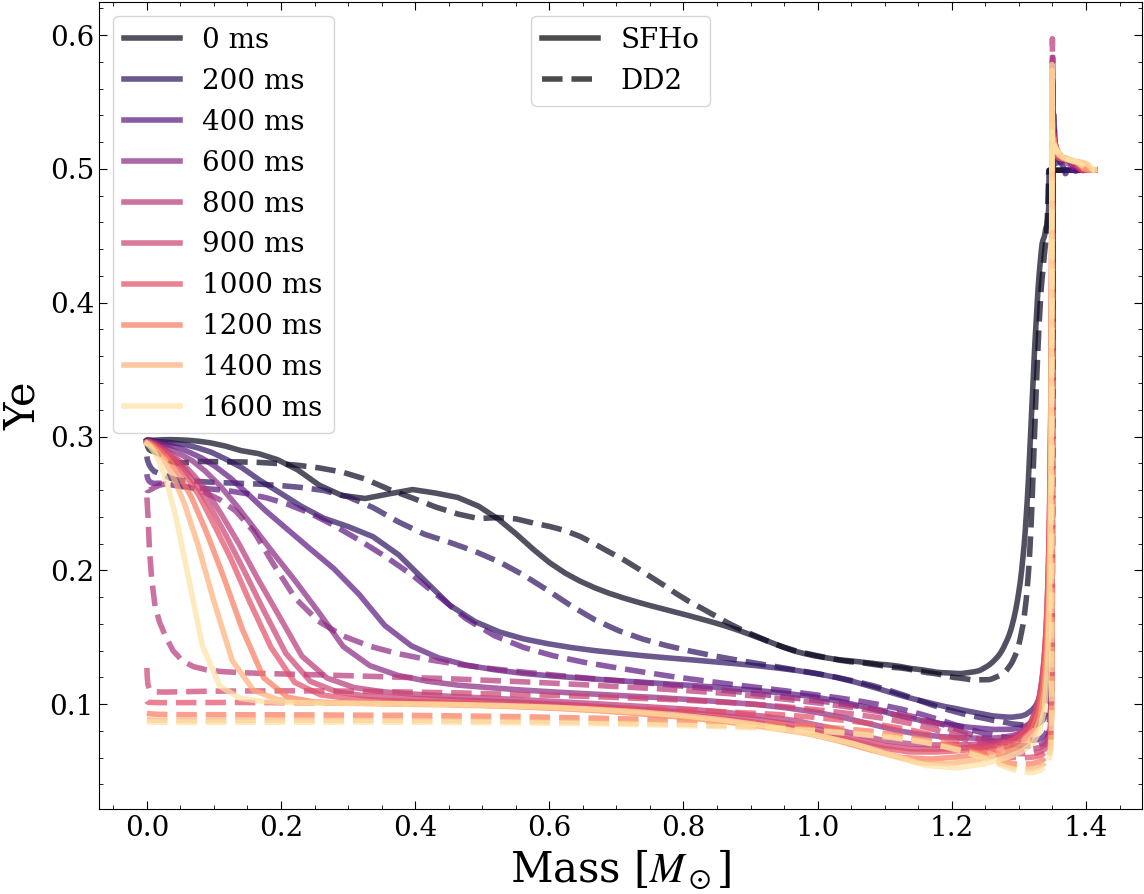}
    \caption{Density (top left), temperature (top right), entropy (bottom left), and electron fraction (bottom right) interior mass profiles in the inner regions for the two EOSes. Time is measured in seconds post bounce and is colored from purple to yellow. See text for a discussion.}
    \label{fig:profiles}      
\end{figure*}

\begin{figure*} 
    \centering
    \includegraphics[width=0.65\textwidth]{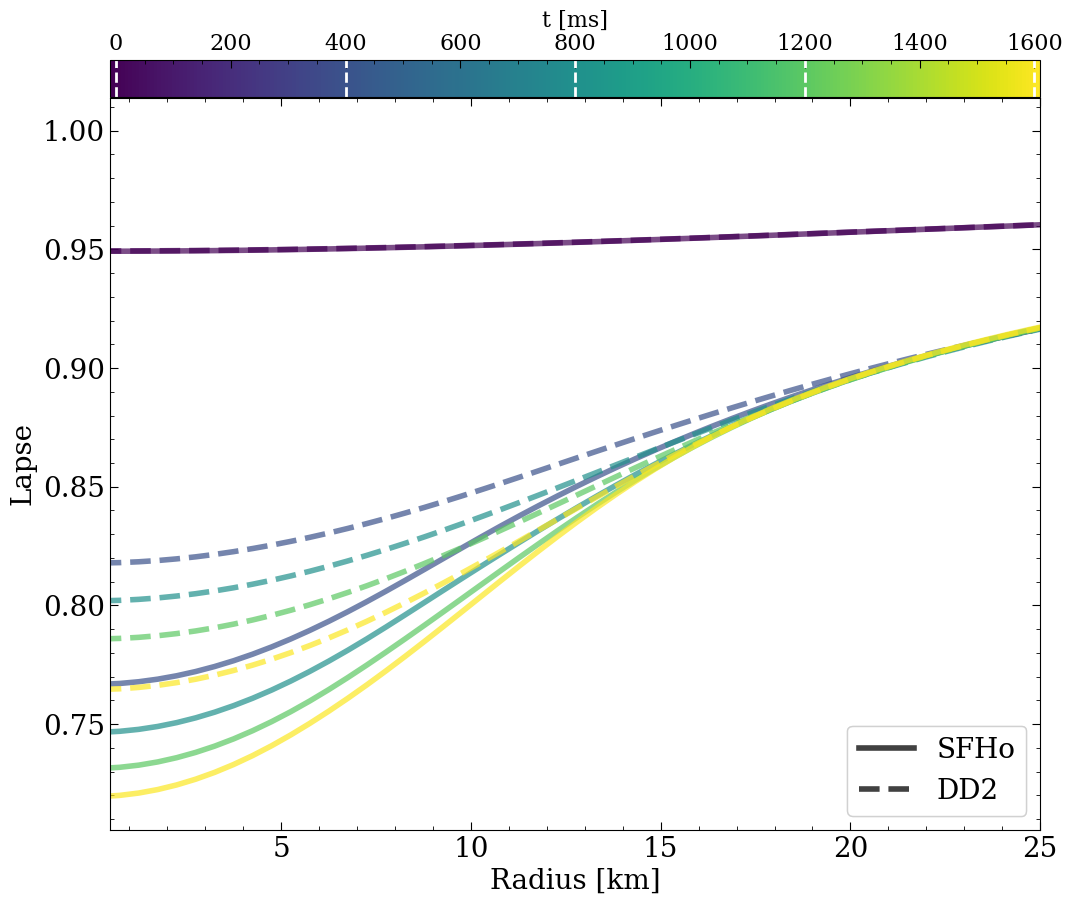}
    \caption{SFHo (solid) and DD2 (dashed) inner lapse profiles versus radius at different times after bounce.}
    \label{fig:lapse}      
\end{figure*}

The panels in Figure \ref{fig:profiles} render the evolution of the average thermodynamic profiles in interior mass of the mass density, temperature, entropy, and Y$_e$ for both equations of state. The final central densities are 5.474$\times$10$^{14}$ g cm$^{-3}$ and 4.485$\times$10$^{14}$ g cm$^{-3}$ for the SFHo and DD2 EOS models, respectively. We note that the densities for the DD2 model are systematically lower. Figure \ref{fig:lapse} shows the lapse profile (redshift from a point to infinity) for various times, again emphasizing the more extended post-bounce configuration for the stiffer DD2 model. The smaller accreted mass and lower resultant PNS mass for a low-compactness 9-$M_{\odot}$ model   
\citep{burrows_correlations_2024} leads to less compression of the shock-generated temperature spike (see Figure \ref{fig:profiles}, top right) by the \adam{weight of the} mass overburden and to smaller temperatures overall. This, along with the smaller densities (and consequently larger radii at a given interior mass) and the associated smaller $\nu_e$ chemical potentials in the electron neutrino trapped region for the DD2 simulation, translates into quicker evolution of the Y$_e$ and entropy profiles \adam{due to faster energy and lepton loss.} This is due in part to the lower resultant \adam{neutrino} ``optical" depths to a given mass shell for the DD2 model. \adam{The lower depths mean that the diffusion time out of the core is shorter for this model.} 

\begin{figure*} 
    \centering
    \includegraphics[width=0.45\textwidth]{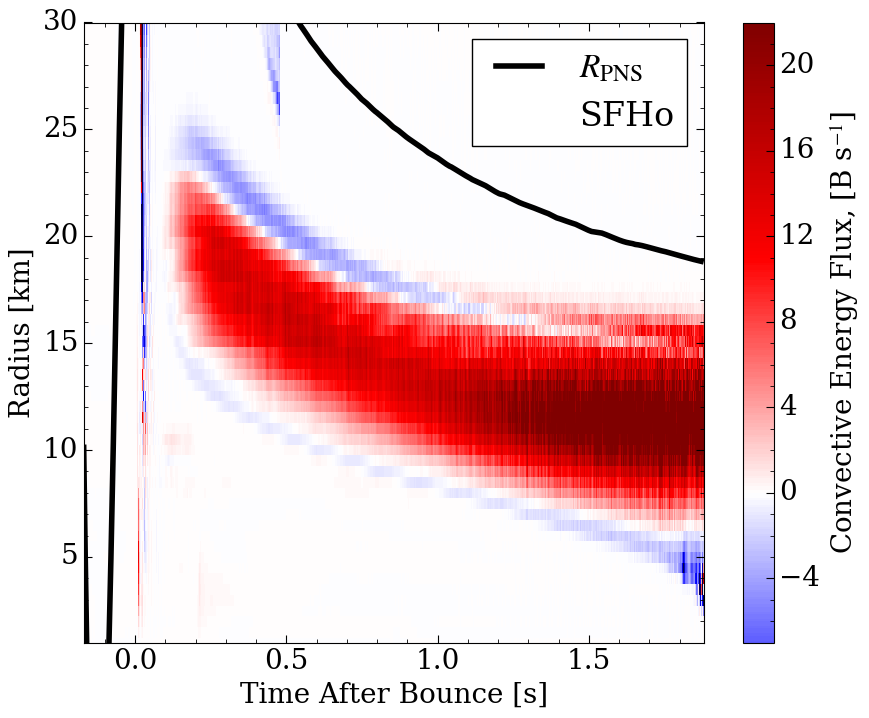}
    \includegraphics[width=0.45\textwidth]{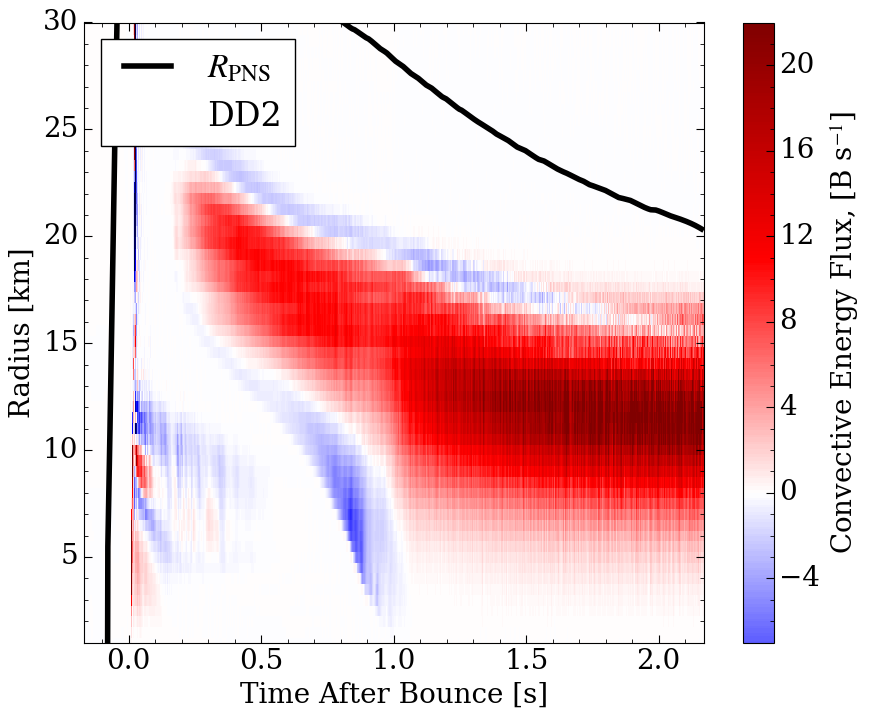}
    \includegraphics[width=0.45\textwidth]{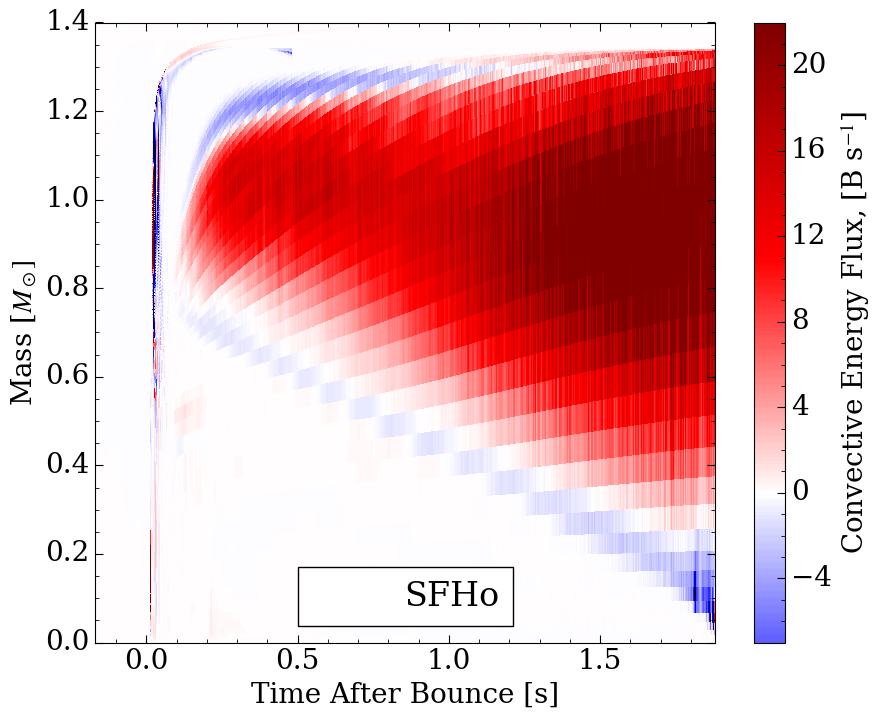}
    \includegraphics[width=0.45\textwidth]{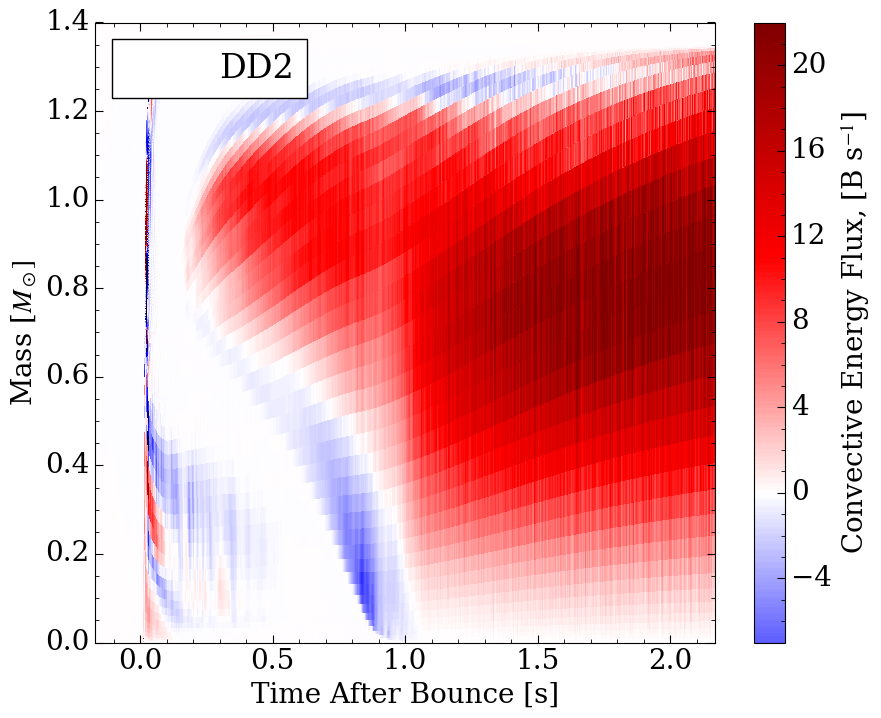}
    \caption{SFHo (left) and DD2 (right) convective flux values at different times after bounce. \adam{The convective flux is calculated using the Bernoulli integral (including the specific internal and kinetic energies, the gravitational potential, and the pressure) and multiplying this by the mean radial mass flux in the convective region.} The top two plots are versus average radius and the bottom two plots are versus interior mass. The black line in the top plots show the evolution of the PNS radius (defined at the mean density isosurface of 10$^{11}$ g cm$^{-3}$). Note that the PNS convective region in the DD2 model achieves the center earlier than seen for the SFHo model. This results in a flattening of the interior Y$_e$ and entropy profiles at earlier post-bounce times (see Figure \ref{fig:shock_r_comp}) than seen using the SFHo EOS and a corresponding boost in the corresponding emergent $\nu_e$ luminosity (see Figure \ref{fig:lum}).}
    \label{fig:conv_flux}      
\end{figure*}

One additional upshot is the earlier onset of penetrating lepton-driven convection that encompasses the whole core sooner for the DD2 model than for the SFHo model. \newadam{Figure \ref{fig:conv_flux} provides a map of the convective flux ($\cal F$), where the red region of the plot, which identifies where PNS convection is significant and positive}, extends to the center within $\sim$1 second of bounce, whereas it does so within $\sim$1.7 seconds for the SFHo model (see also \citet{Nagakura2020}), for which the core is more compact. 

\newadam{Here, we calculate the convective flux ($\cal{F}$) using the Bernoulli integral ($\cal{B}$) (including the specific internal ($\varepsilon$) and kinetic ($\frac{1}{2}v^2$) energies, the gravitational potential ($\phi$), and the pressure ($P$)) and multiplying this by the net radial mass flux relative to the angular mean at each radius in the convective region. Specifically, 

  \begin{equation}     
        {\cal F} = \langle \rho (v_{\rm r}-\langle v_{\rm r}\rangle){\cal{B}}\rangle  = \langle\rho (v_{\rm r}-\langle v_{\rm r}\rangle)(\varepsilon + P + \frac{1}{2}\ v^2 + \phi)\rangle\,,
\end{equation}
where $\langle A\rangle$ is the average over the sphere.
}

\begin{figure*} 
    \centering
    \includegraphics[width=0.65\textwidth]{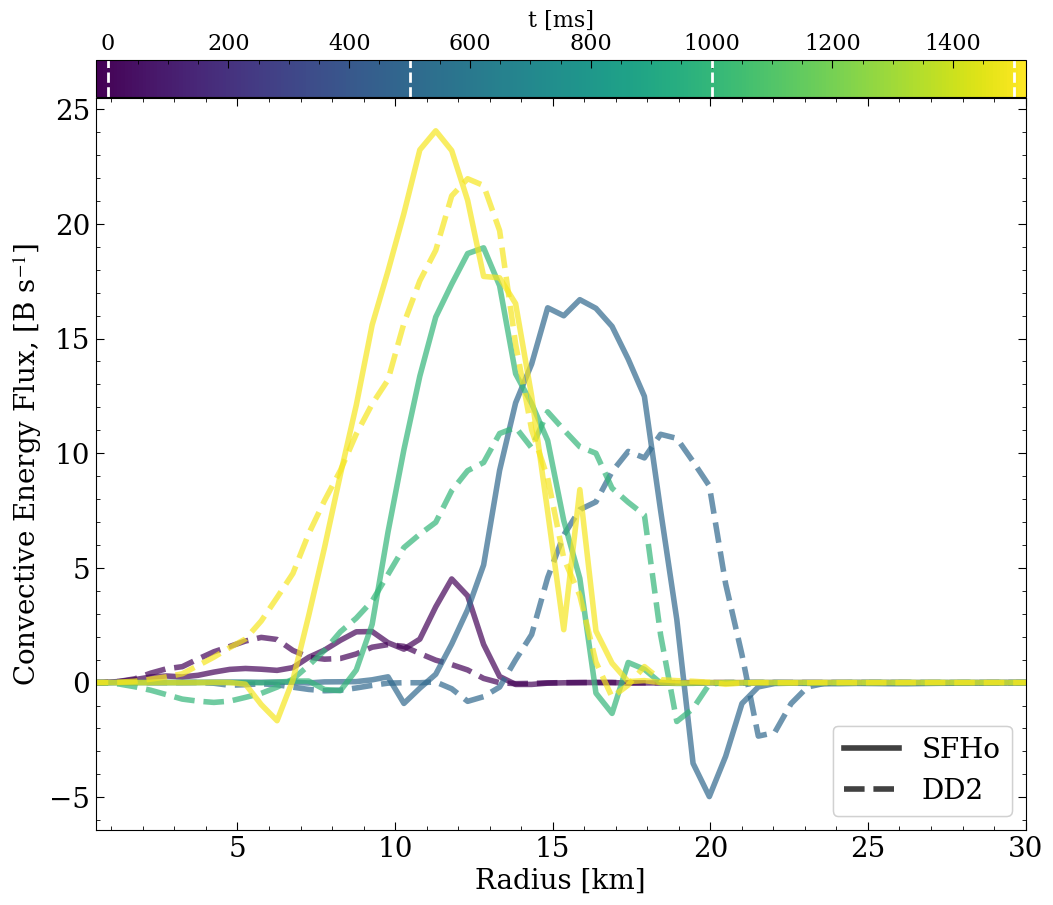}
    \caption{Angle-averaged convective flux versus radius (in km) in the proto-neutron star (PNS) for various times after bounce (color bar). Note that early on the radius range experiencing PNS convection for the DD2 run is exterior to that of the SFHo run.  However, later on near $\sim$1 second after bounce, its inner boundary extends interior to that of the SFHo run.  Moreover, the peak convective flux for the DD2 run starts below that of the SFHo run by $\sim$$\frac{1}{3}$, but later near $\sim$1 second after bo   unce begins to compete with that for the SFHo run and (as Figure \ref{fig:conv_flux} demonstrates) achieves the inner zones to encompass most of the PNS at a significantly earlier time. See text for a discussion.}
    \label{fig:conv_flux2}      
\end{figure*}

The differences in the mean convective luminosity profile for various times after bounce and for the two EOS models is given in Figure \ref{fig:conv_flux2}. At later times, the dashed curves representing the DD2 EOS extend into the deep interior at an earlier phase around $\sim$1 second after bounce. This early onset of significant convective transport to the surface (and then to infinity) is one interesting characteristic of the DD2 model.

\section{Differences in the Luminosities and Mean Energies}
\label{lum}

\begin{figure*}   
    \centering
    \includegraphics[width=0.65\textwidth]{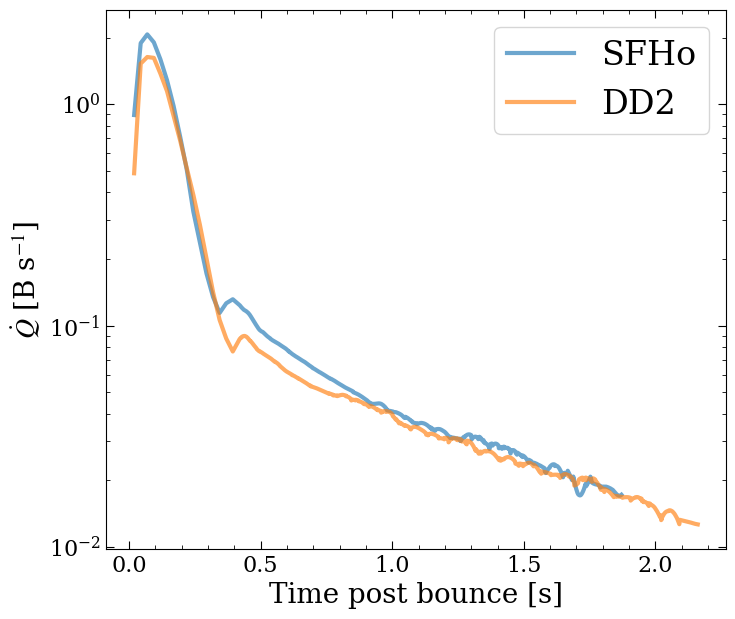}
    \caption{$\dot{Q}$ (the net heating rate in the gain region) versus time after bounce for the two equations of state. Note that during the early shock launching and driving phase the heating rate for the DD2 EOS is lower.  This translates into the lower asymptotic explosion energy for this EOS seen in Figure \ref{fig:energy}.}
    \label{fig:Qdot_comp}      
\end{figure*}

\begin{figure*}   
    \centering
    \includegraphics[width=0.65\textwidth]{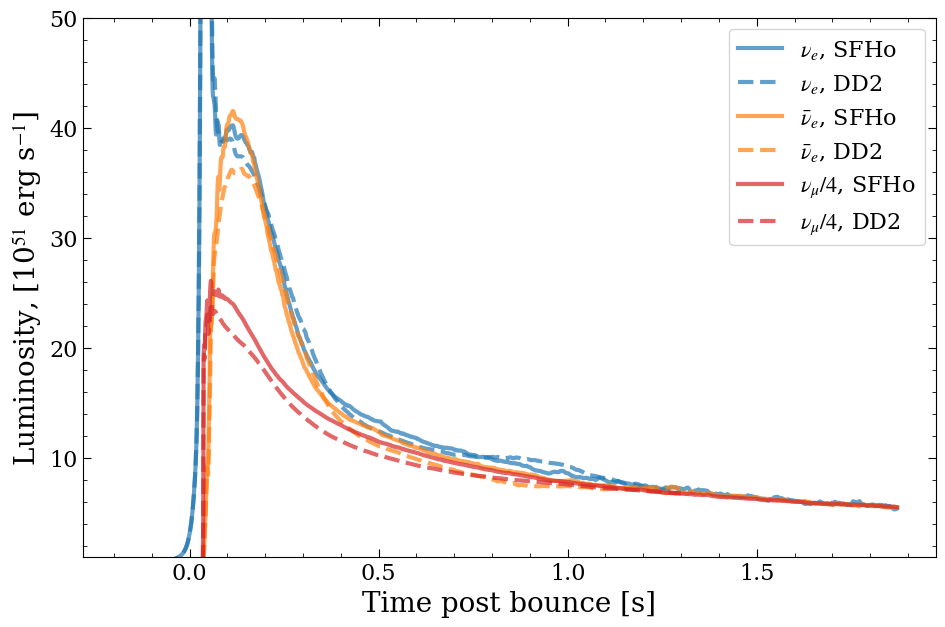}
    \includegraphics[width=0.65\textwidth]{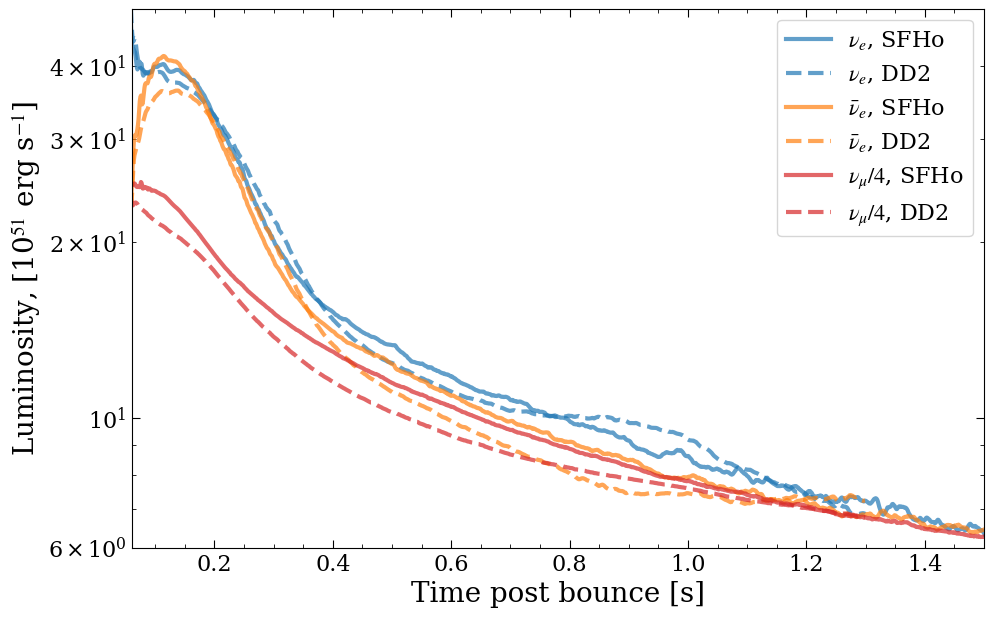}

    \caption{Neutrino luminosities (top) for the various neutrino types as a function of time post bounce (in seconds) for the SFHo (solid) and DD2 (dashed) EOSes. The ``$\nu_{\mu}$/4" lines are for the ``$\nu_{\mu}$"'s singly. Note that the data for this plot were dumped a bit coarsely and, hence, missed the pre-bounce precursor bump in the $\nu_e$ neutrino luminosity curves, while capturing the breakout burst. Note, as the bottom plot emphasizes, that at late times around one second the $\nu_e$ neutrino luminosity for the DD2 EOS actually exceeds that for the SFHo EOS.  This is due to the earlier onset of extensive PNS convection in the DD2 context.  See text for a discussion.}
    \label{fig:lum}      
\end{figure*}

\begin{figure*}   
    \centering
    \includegraphics[width=0.65\textwidth]{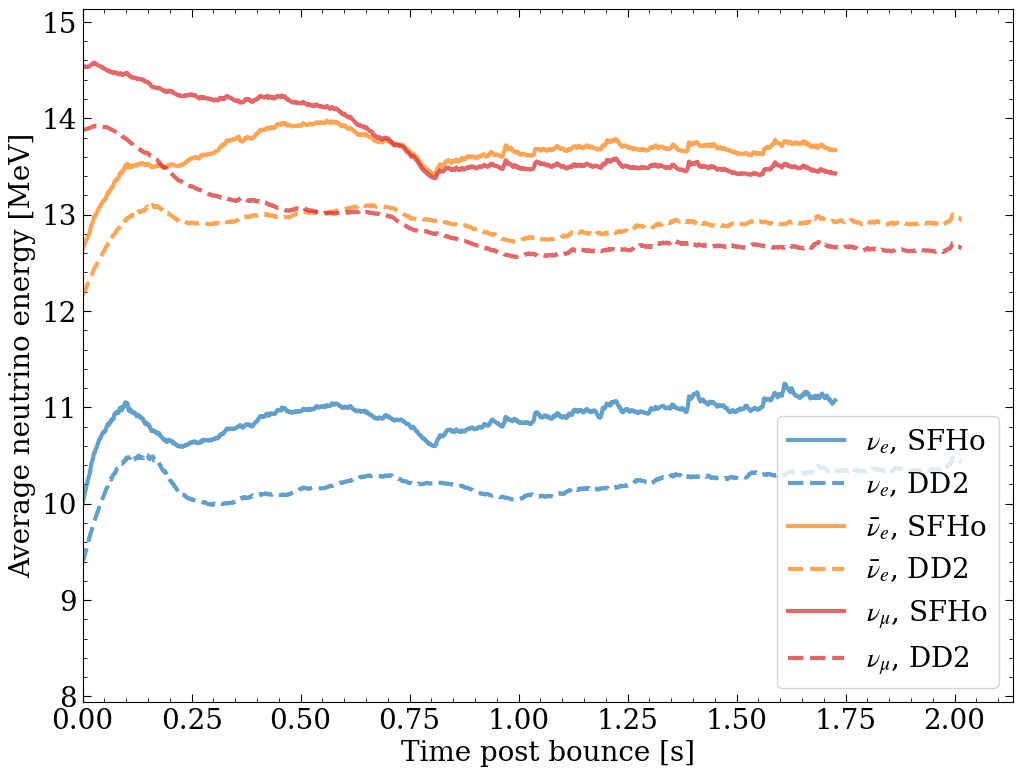}
    \caption{Average neutrino energy (in MeV) for the three neutrino species ($\nu_e$ (blue), $\bar{\nu}_e$ (orange), and ``$\nu_{\mu}$" (red) for the SFHo (solid) and DD2 (dashed) equations of state versus time after bounce. Note that the mean neutrino energies are consistently lower for the DD2 model.}
    \label{fig:average_energy}      
\end{figure*}

The neutrino luminosity evolution for each neutrino species is portrayed in Figure \ref{fig:lum}. We note that for the first $\sim$0.2 seconds after bounce, the phase during which the shock is energized by neutrino heating (see Figure \ref{fig:Qdot_comp}), the $\nu_e$ and $\bar{\nu}_e$ luminosities are lower for the DD2.  As important as we see in Figure \ref{fig:average_energy}, the mean neutrino energies are systematically lower for the DD2 model.  Since the absorption cross sections of electron neutrinos on the nucleons in the gain region are increasing functions of neutrino energy, lower mean $\nu_e$ and $\bar{\nu}_e$ neutrino energies and lower corresponding luminosities result in the lower neutrino heating rates seen in Figure \ref{fig:Qdot_comp}.  These are the main reasons for the differences seen between the SHFo and DD2 models and is as expected (\S\ref{sec:int}). However, we have here quantified them.

It is interesting to point out that the time of the penetration of PNS convection deep into the core in the DD2 model (at $\sim$1 second after bounce) coincides with a $\sim$10-15\% enhancement in the $\nu_e$ luminosity, and a $\sim$5\% decrease in the $\bar{\nu}_e$ luminosity. This is most easily seen in the bottom panel of Figure \ref{fig:lum}. Though a modest difference, such behavior is to be expected and may be a weak diagnostic of the DD2 model and a distinguishing characteristic of its longer-term cooling. However, whether this difference and its earlier manifestation in the DD2 model could be discernible in subsurface neutrino detectors on Earth remains to be determined.


\section{Explosion Energy and Kicks}
\label{energy}

The final gravitational mass of the residues of collapse for the two models are 1.242 $M_{\odot}$ and 1.259 $M_{\odot}$ for the SFHo and DD2, respectively (see Table \ref{tab:main}). The corresponding final gravitational mass for the same progenitor (model 9b) published in \citet{burrows_correlations_2024} (using the SFHo EOS) was 1.238 $M_{\odot}$. \adam{The very slight difference between the residual mass of the current and the previous (9b) 9-$M_{\odot}$ model is likely due to the chaos in turbulent flow, which is both numerical and physical.} \adam{The greater residual mass of the DD2 model reflects the longer delay to explosion and the greater degree of subsequent accretion for its weaker explosion. \footnote{\adam{Further discussion of the features of models 9a and 9b can be found in \S\ref{nucleo}.}}}

\begin{figure*}   
    \centering
    \includegraphics[width=0.65\textwidth]{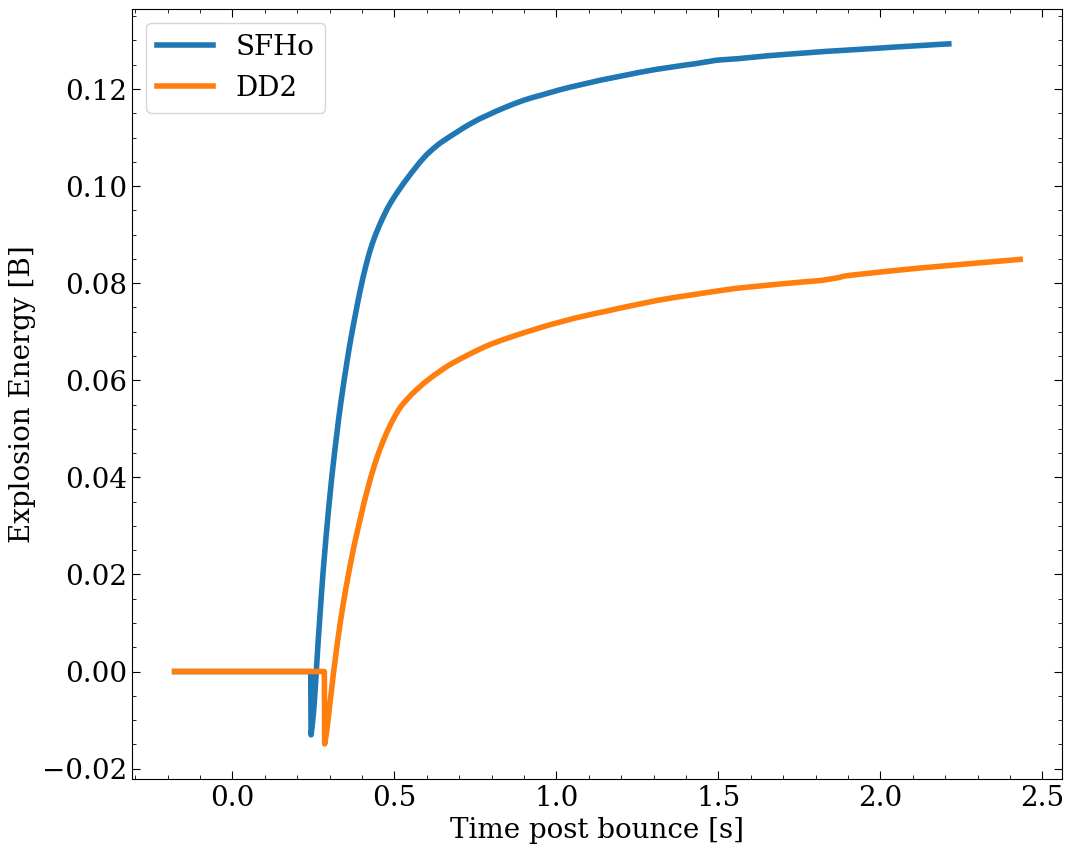}
    \caption{Estimate of the total explosion energy (thermal plus kinetic plus gravitational plus off-grid envelope sum) versus time as it asymptotes to its final value. Note that it starts negative after the onset of explosion due to the net bound character of the matter that will eventually be ejected. Ongoing neutrino heating after shock launching is required to achieve the final (positive) explosion energy.}
    \label{fig:energy}      
\end{figure*}

One of the most interesting differences between these two EOS models is the asymptotic explosion energy.  \newadam{The explosion energy includes the kinetic, thermal, and gravitational energies of the ejected matter. Specifically, it is calculated as the sum of $\varepsilon + \frac{1}{2}v^2 + \phi$ over all the matter with a positive value of this quantity exterior to 1000 km. The recombination energy and binding energy of the overburden mantle outside the computational grid are also included. At the end of the simulation this is the actual explosion energy at infinity.} This is not the ``diagnostic" energy sometimes quoted, but the total final energy at ``infinity," including the binding energy of the outer mantle not on the computational grid. Figure \ref{fig:energy} shows its development to the end of the simulations, at which point its residual change is quite small. Note that the energy starts negative due to the negative mantle binding energy but soon becomes positive due to neutrino heating after the onset of the explosion. The neutrino heating before the explosion is irrelevant to the final explosion energy \citep{burrows:95}. We see that the explosion energy for the SFHo EOS ($\sim$0.123 B [$\equiv$ 10$^{51}$ ergs) is $\sim$50\% larger than that for the DD2 EOS ($\sim$0.08 B)\footnote{Curiously, the explosion energies scale very approximately as the square of the final central density. The direction of the scaling makes sense, but we can as yet discern no clear physical argument for such a relationship.}. 

We comment that, \adam{unlike the 9a and 9b models published in \citet{burrows_correlations_2024}}, the models of this paper were collapsed in 3D (not 1D).  Given this, small numerical perturbations were allowed more time to grow during collapse than in the previous \adam{model 9b of \citet{burrows_correlations_2024}), making it more like model 9a of that paper for which perturbations were imposed.} The upshot is slightly larger seeds post-bounce for the neutrino-driven convective instability, resulting in a quicker explosion. \adam{The physical reason for the difference lies in the fact that larger seed perturbations jumpstart the growth of the turbulence which, through the associated quicker rise in the turbulent pressure behind the shock and the slight earlier onset of PNS convection in the core, facilitates the launch of the stalled shock wave. This general component of CCSN theory is described in \citet{wang,2021Natur.589...29B} and \citet{burrows_correlations_2024}.} The magnitude of these seed perturbations in the models of this paper in fractional mass density interior to the stalled shock just $\sim$1 millisecond (ms) after bounce was $\sim$0.9\% RMS and 3\% maximum. \adam{The corresponding magnitude of the seeds for the legacy 9b model after $\sim$20 milliseconds after bounce was $\sim$0.1\%.} This is the origin of the \adam{slight difference in the explosion energy between our SFHo model here and the SFHo model 9b} published in \citet{burrows_correlations_2024} (0.095 B [$\equiv\ 10^{51}$ ergs]) and why it resembles that of the 9a model (0.111 B), for which initial seed perturbations were imposed.  \adam{However, these differences are significantly smaller than the difference in explosion energy we find between the SFHo and DD2 models.}

It is worth commenting that the fractional increment in explosion energy is larger than the fractional increment in the neutrino heating rate (Figure \ref{fig:Qdot_comp}). This non-linearity is a common feature of CCSN simulations and emphasizes the sensitivity of the asymptotic state to the input physics \citep{burrows_correlations_2024,boccioli2024}.   

\begin{figure*}   
    \centering
    \includegraphics[width=0.75\textwidth]{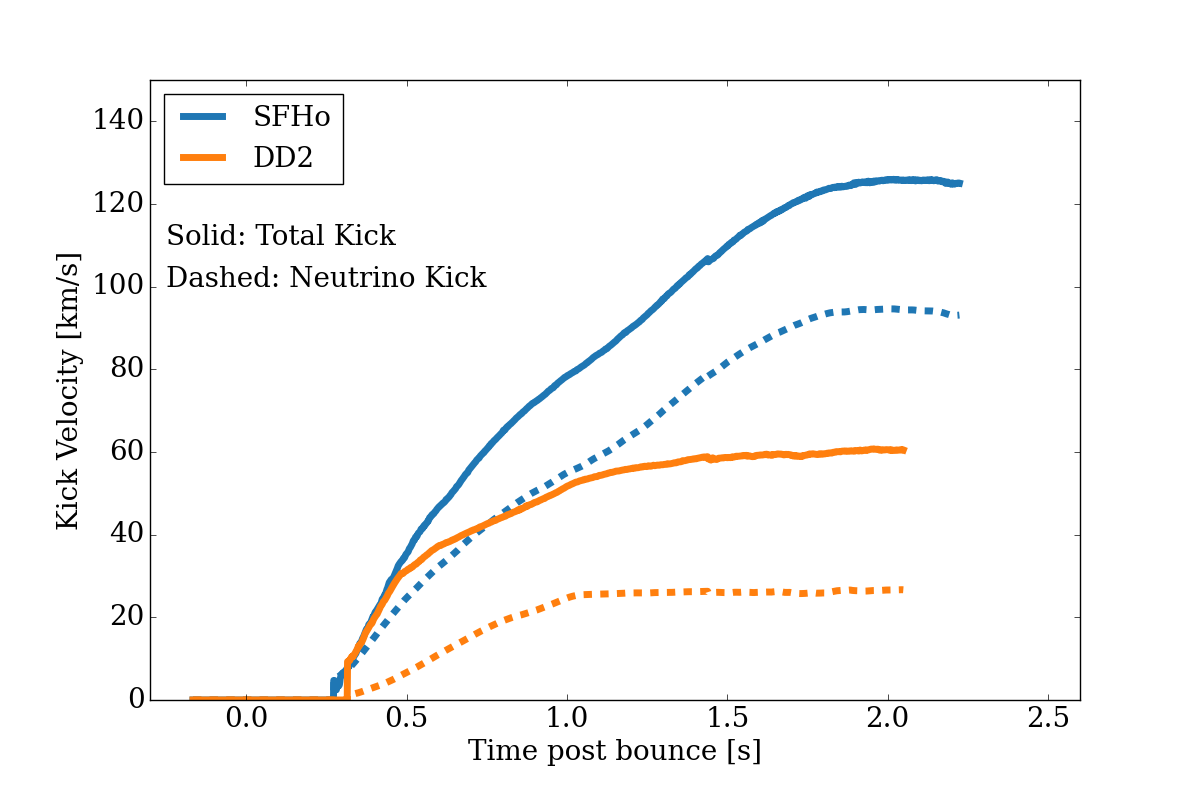}
    \caption{Absolute value of the kick velocities versus time for the SFHo (blue) and DD2 (orange) equations of state. \adam{Solid lines shows the total kick magnitude, while the dashed lines show the magnitude of the neutrino kick component. Note that these two components are not in general aligned and must be added vectorially.}}
    \label{fig:kick}      
\end{figure*}

Figure \ref{fig:kick} displays the evolution of the recoil kicks experienced by the newly-born neutron star for the two EOS runs. In general, there is some degree of stochasticity in the final direction and magnitude of the kicks due to the chaos in the turbulent flow.  In particular, the direction is difficult to predict when there is no initial spin. However, the final kick speeds are $\sim$132 km s$^{-1}$ and $\sim$68 km s$^{-1}$ for the SFHo and DD2 runs, respectively \footnote{These numbers include the net vector kick due to neutrino recoil (see Figure \ref{fig:kick}), which for low-compactness models can be a substantial fraction. \adam{In particular, the magnitude of the final net neutrino kick for the SFHo and DD2 models is $\sim$98 and $\sim$26 km s$^{-1}$, respectively.  However, the directions do not align with the net matter recoil and the resultant kick is a vector sum. This general point is discussed in detail in \citet{burrows_kick_2024}.}}. The former was what was found for our previous 9-$M_{\odot}$ simulation (9a) \citep{burrows_correlations_2024} ($\sim$120.7 km s$^{-1}$), while the lower value of the latter is what might be expected for a lower energy explosion, ceteris paribus \citep{Janka2017}. One is tempted to conclude that this difference might be universal across the exploding-model continuum, but such a conclusion is premature in lieu of a more substantial set of progenitor simulations.


\section{Nucleosynthesis}
\label{nucleo}

\begin{figure*}   
    \centering
    \includegraphics[width=0.47\textwidth]{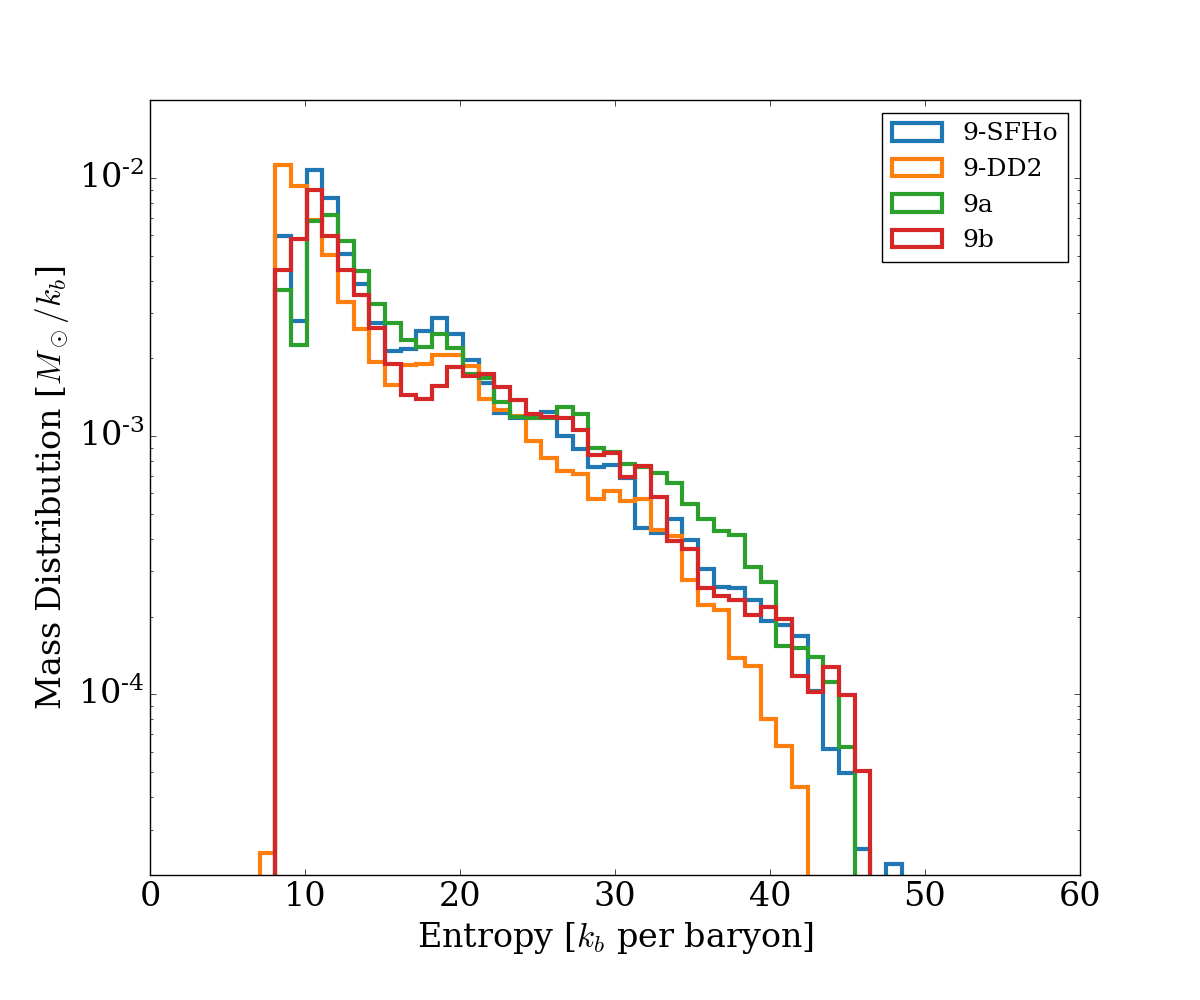}
    \includegraphics[width=0.47\textwidth]{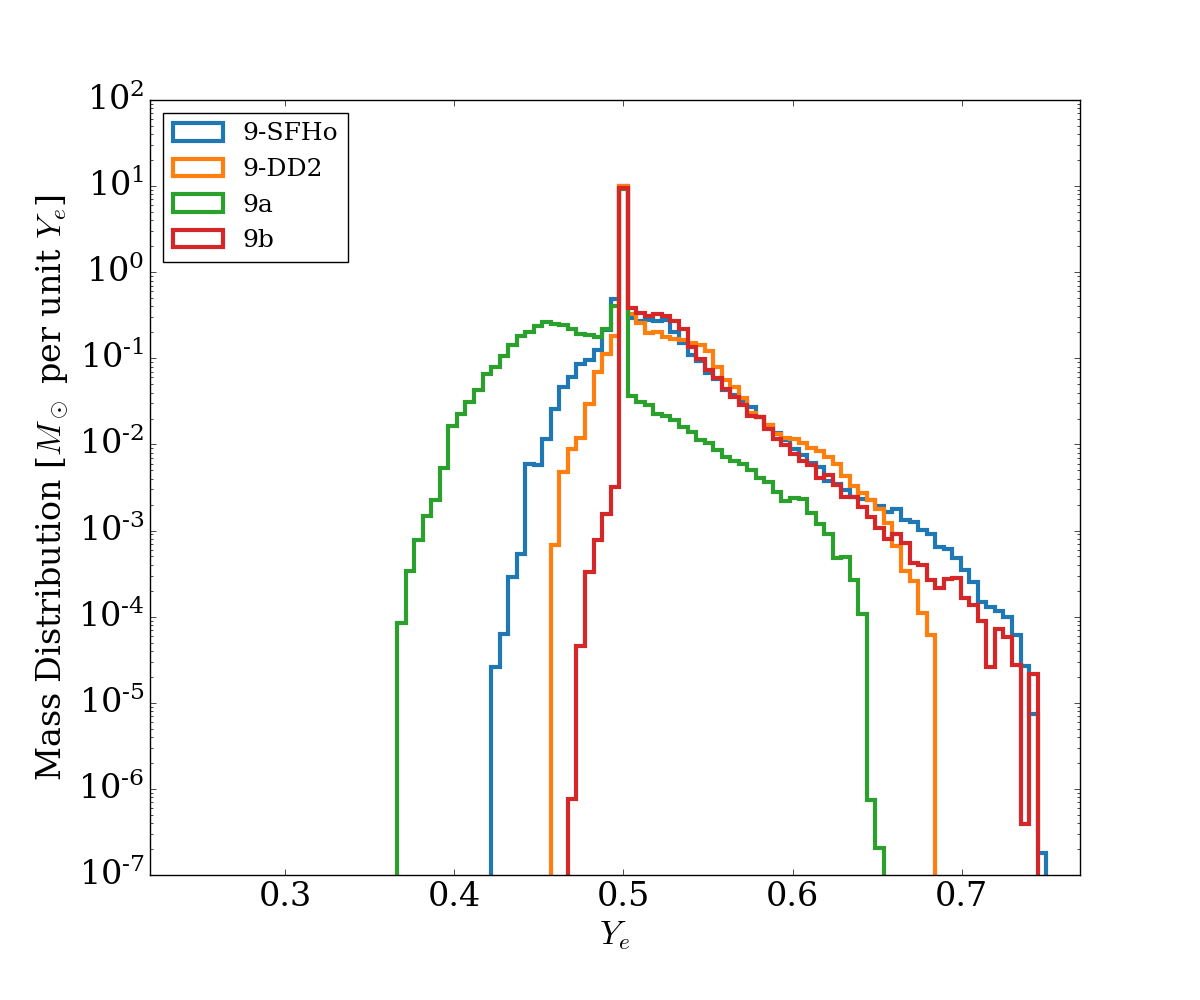}
    \caption{The ejecta entropy and electron fraction distributions for the SFHo (blue) and DD2 (orange) equations of state models. Two other 9$M_\odot$ models using the SFHo EOS from \citet{wang_nucleo_2024}, 9a (green) and 9b (red), are shown for comparison. The differences between the 9-SFHo, 9a, and 9b models can be explained by their different treatment of the initial perturbations. See main text for a more detailed explanation. While the entropy distributions look almost identical, the 9-SFHo model has more neutron-rich ejecta than the 9-DD2 model because of the earlier and more energetic explosion. However, this effect is significantly weaker than that of initial perturbations due to turbulence in the progenitor burning regions.}
    \label{fig:S-ye}      
\end{figure*}

\begin{figure*}   
    \centering
    \includegraphics[width=0.47\textwidth]{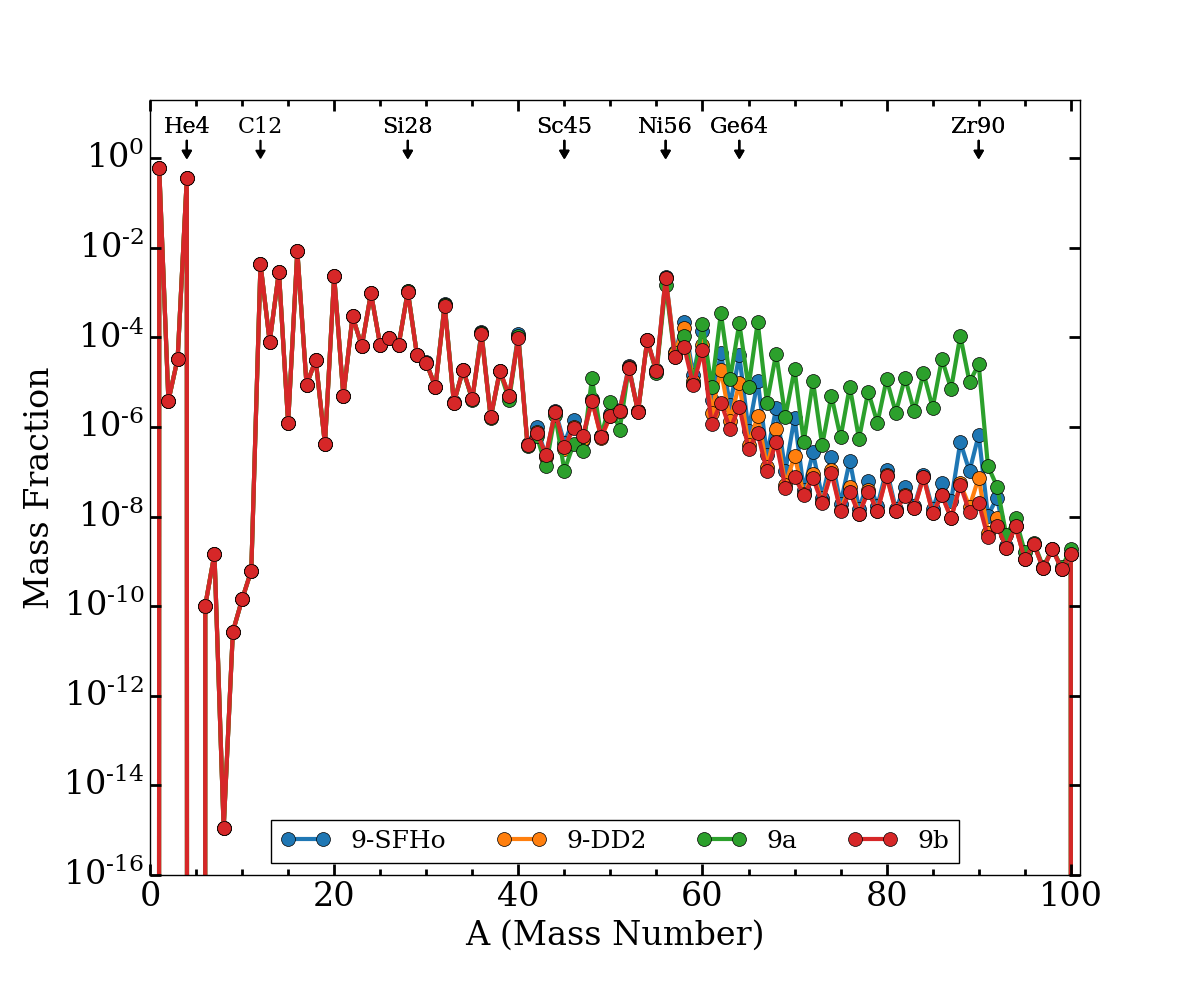}
    \includegraphics[width=0.47\textwidth]{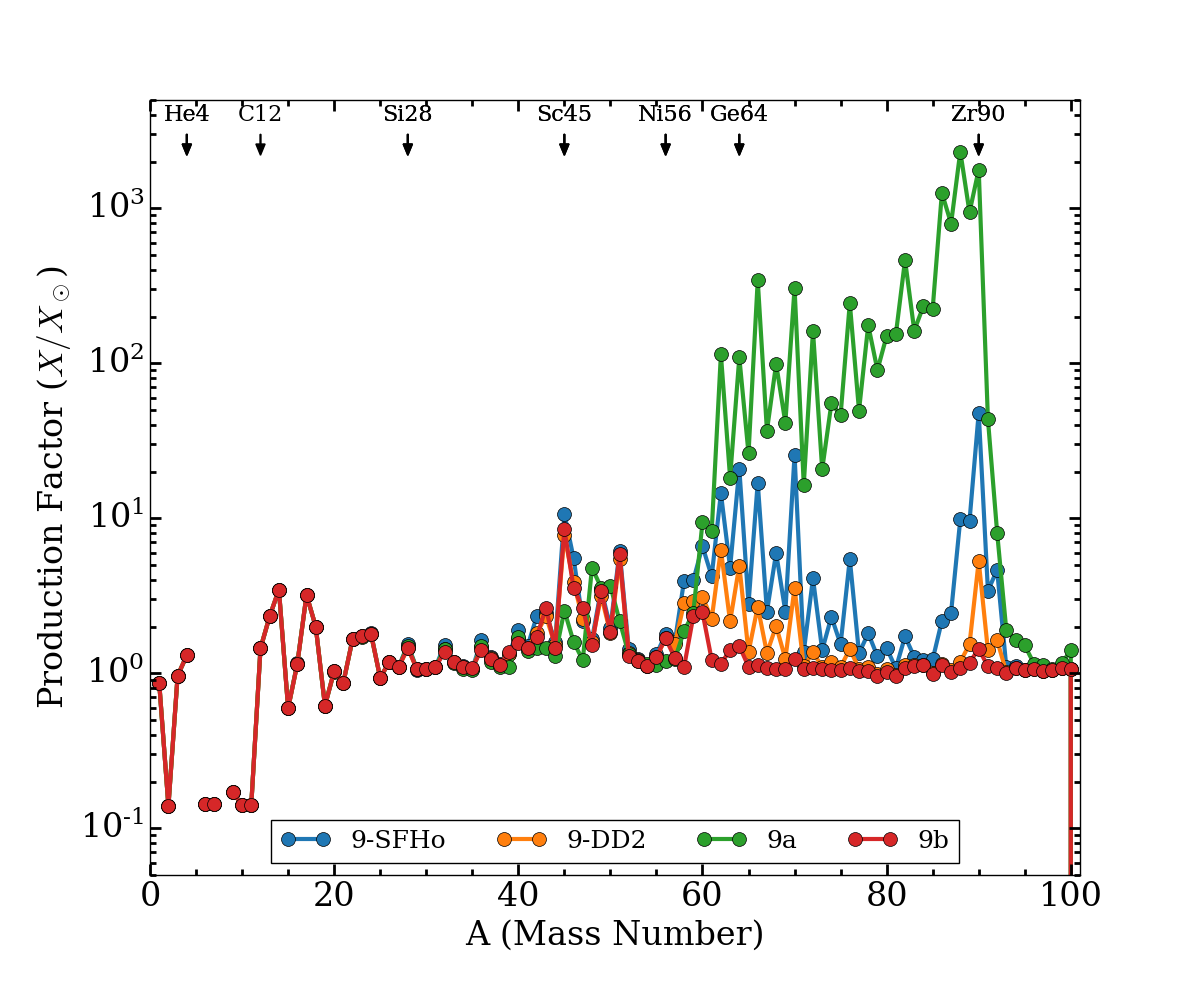}
    \caption{The ejecta mass fractions and production factors at the end of the SFHo (blue) and DD2 (orange) model calculations. Two other 9$M_\odot$ models using the SFHo EOS from \citet{wang_nucleo_2024}, 9a (green) and 9b (red), are shown for comparison. The differences between the 9-SFHo, 9a, and 9b models can be explained by their different treatment of initial perturbations. The abundance differences are a direct consequence of the $Y_e$ differences shown in Figure \ref{fig:S-ye}. The 9-SFHo model shows more heavy elements and a more enhanced $^{90}$Zr peak than the 9-DD2 model, but the EOS effect is significantly weaker than that of an initial perturbation, as exemplified in model 9a.}
    \label{fig:yield}      
\end{figure*}

Figure \ref{fig:S-ye} shows the asymptotic entropy and electron fraction distributions of the ejected matter in the SFHo and DD2 models of this paper. For comparison, two other 9$M_\odot$ models using the SFHo EOS from \citet{wang_nucleo_2024} and \citet{burrows_correlations_2024}, 9a and 9b, are also shown. The differences between the 9-SFHo, 9a, and 9b models can be explained by their different treatment of initial perturbations. Both the 9a and 9b models were evolved spherically (1D) until 10 ms post-bounce and continued in 3D after that. To the radial velocity field of the 9a model 10 ms after bounce, we artificially added an $l=12$ spherical harmonic perturbation field between 200 and 1000 km with a maximum $\Delta v$ of 100 km s$^{-1}$, using the prescription of \citet{muller_janka_pert}, while the 9b model was unmodified. However, the collapse and bounce phase of the 9-SFHo and 9-DD2 models were performed in full 3D. Although the collapse phase shows no difference between 1D and 3D, the 10 ms post-bounce evolution period evolves differently. During this interval, the 9-SFHo and 9-DD2 were able to follow the development of so-called ``prompt" convection in the shocked mantle of the proto-neutron star. Therefore, at 10 ms post-bounce, the 9b model is spherically symmetric, the 9a model has velocity perturbations between 200 and 1000 km mimicking perturbations in the convective oxygen burning shell, and the 9-SFHo and 9-DD2 models have early velocity perturbations in the PNS due to  ``prompt" convection. The different early perturbation treatments lead to slight differences in explosion times and energies, but also affect the asymptotic $Y_e$ distributions due to the differential degrees of explosive expansion and $Y_e$ freeze-out \citep{pruet2006}.  While the asymptotic ejecta entropy distributions look almost identical, the 9-SFHo model has more neutron-rich ejecta than the 9-DD2 model because of its earlier and more energetic explosion, which leads to slightly more neutron-rich ejecta. \adam{However, as Figure \ref{fig:yield} demonstrates around an atomic weight ($A$) of 80,} this EOS effect is significantly weaker than that seen in model 9a with an imposed initial perturbation \adam{that proved larger than that for any of the other models 9b, 9-SFHo, and 9-DD2.}. 

For our nucleosynthetic calculations, we use the Skynet code \citep{lippuner2017} and the approach to its use outlined in \citet{wang_nucleo_2024}. \adam{As described there, we identify  at the end of the simulation $\sim$321,000 ejecta parcels, integrate them backwards to establish the approximate Lagrangian trajectories of each, and then post-process each parcel individually along its trajectory with Skynet, following $\sim$1560 isotopes. Then, we sum the individual yields of the ensemble to establish the integral yields.} Figure \ref{fig:yield} displays the mass abundances and production factors of the models. Curiously, the SFHo and DD2 models have similar $^{56}$Ni and $^{44}$Ti yields (within $\sim$15-20\%, see also Table \ref{tab:main}). Though the DD2 model has a lower explosion energy, its ejecta are also less neutron-rich, which by itself would tend to enhance $^{44}$Ti and $^{56}$Ni production.  The countervailing combination of explosion energy and neutron-richness of the ejecta for the two EOSes results in similar yields for both $^{56}$Ni and $^{44}$Ti. Also, the faster-exploding 9-SFHo model shows more heavy elements and a more enhanced $^{90}$Zr peak than the 9-DD2 model.  \adam{As noted above,} in Figure \ref{fig:S-ye} we see for the SFHo models (9b, 9-SFHo, 9a) the progressive increase in the contribution of neutron-rich ejecta parcels with degree of early post-bounce perturbation. This is reflected in the hierarchy of yields and production factors in the $A=60-90$ regions depicted in Figure \ref{fig:yield} and suggests that such perturbations can enhance weak r-process \citep{woosley2002,Travaglio2004,Cowan2021,wang_nucleo_2024,Okada.weak.rprocess.2026} production for faster-exploding models. This is what we see in the comparison of yields for the SFHo and DD2 models, where the SFHo model exploded a bit more quickly and energetically. As a consequence, it has more neutron-rich ejecta parcels and slightly greater yields from $A=60$ to 90. The corresponding yields for the 9a and 9b models are also given in \citet{wang_nucleo_2024}. 

We note that in the past, for the same EOS, we did not see a dependence on the magnitude of seed perturbations for models that exploded later \citep[e.g., a 18.5 $M_{\odot}$ progenitor,][]{wang_wind,wang_nucleo_2024}, ostensibly due to the fact that if explosion occurs after the post-shock turbulence is fully saturated the role of initial perturbation is muted. This suggests an important role for the lower progenitor mass, faster-exploding CCSNe in the production of the weak r-process, among other nucleosynthetic components that are generated in low $Y_e$ environments. However, it will clearly be important to conduct an extended EOS comparison study for a broad range of progenitor masses, equations of state, and compactnesses.



\section{Gravitational Radiation}
\label{gw}

\begin{figure*}   
    \centering
    \includegraphics[width=0.9\textwidth]{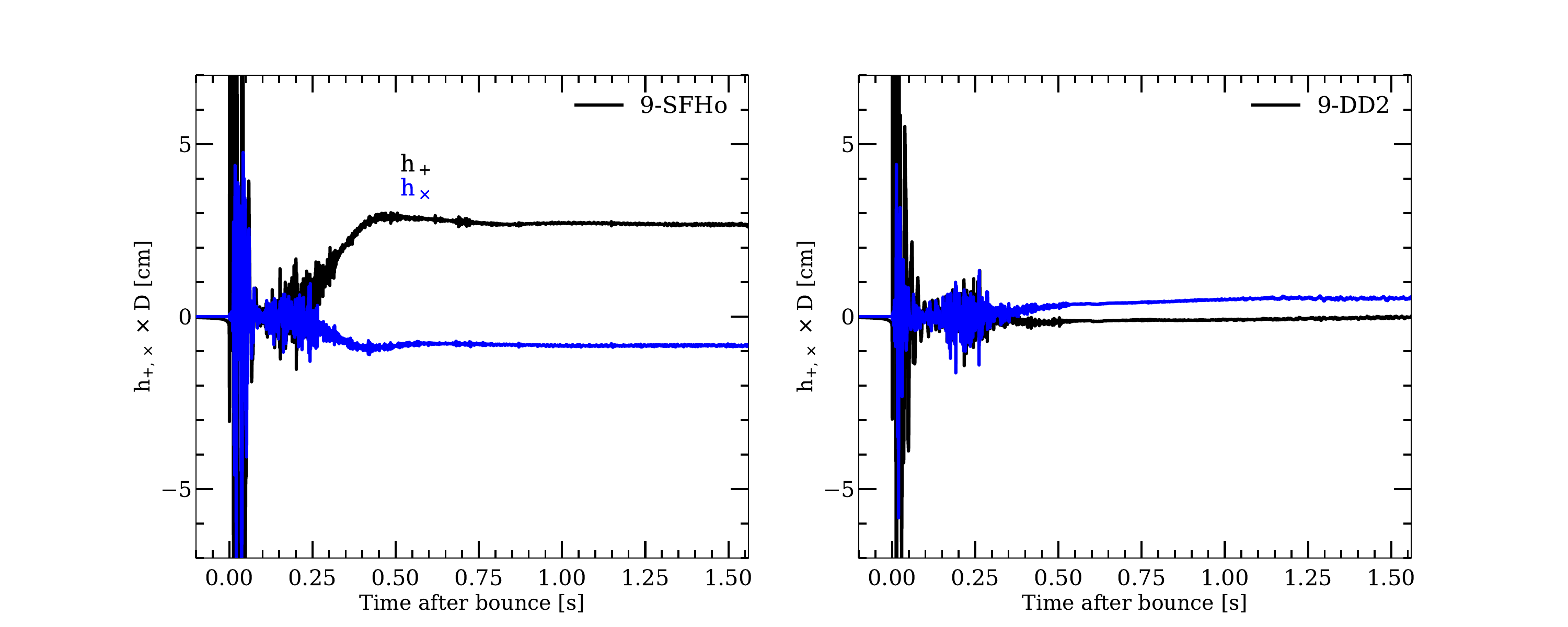}
    \includegraphics[width=0.47\textwidth]{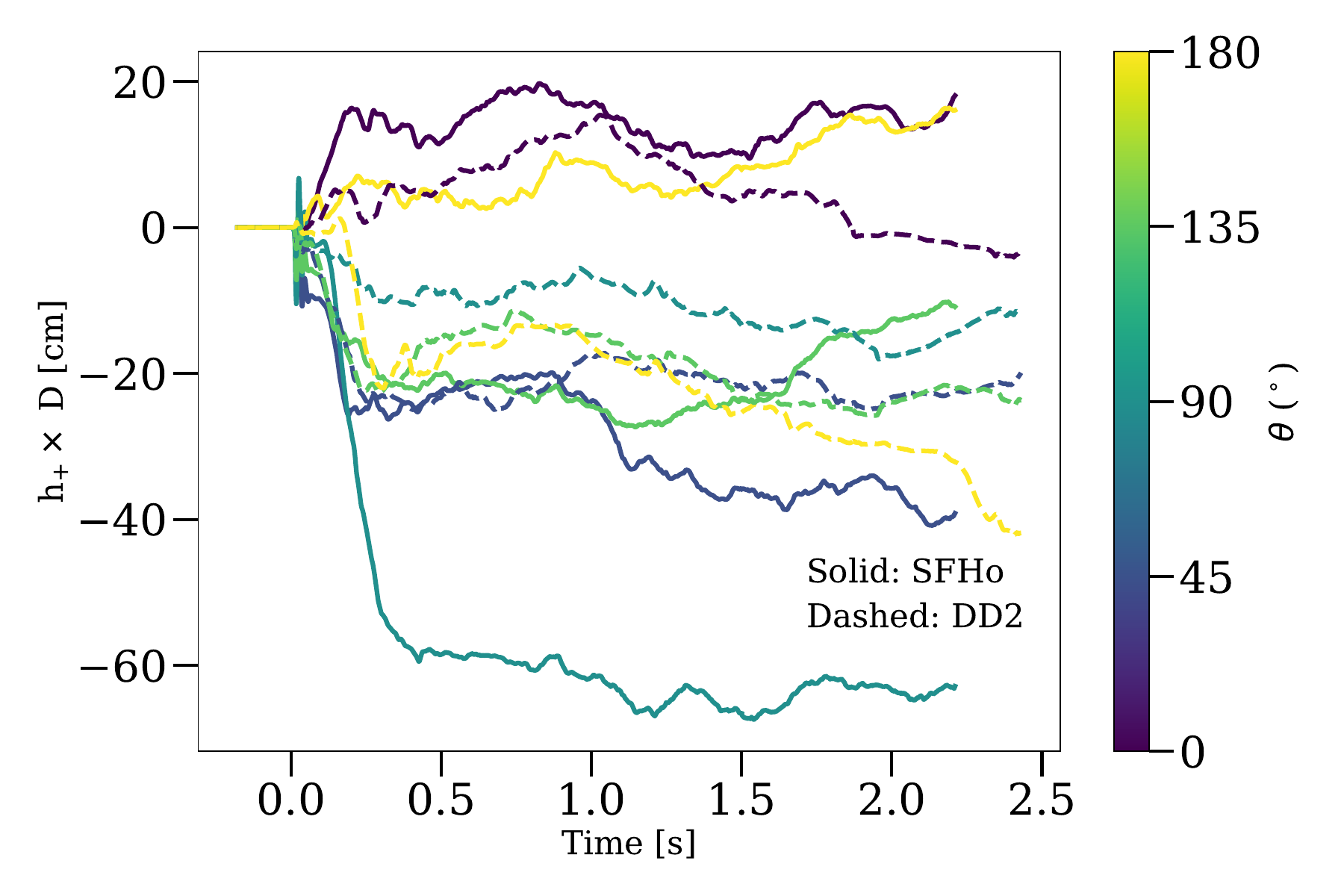}
    \caption{Comparison of the matter gravitational-wave strains (\textbf{top}) for the SFHo (left) and DD2 (right) equations of state and the corresponding neutrino memory strains (\textbf{bottom}) for the $x, \times$ polarization \adam{along a sample of $\theta$ directions}. After the first 500 ms, due to the large asymmetry in its explosion, the SFHo EOS results in higher matter and neutrino memory strains. See text for a discussion.}
    \label{fig:strain}      
\end{figure*}

Figure \ref{fig:strain} portrays the gravitational wave strain along the x-axis times the distance to the supernova for the SFHo (top left) and DD2 (top right) simulations, for both the $+$ (black) and $\times$ (blue) polarizations. The salient aspects of this signature are the early burst in the first $\sim$50 milliseconds after bounce, followed by the stochastic signature of pelting accretion excitation of the PNS \citep{vartanyan2020}, with the matter asymmetry memory component \citep{choi2024} rounding out the longer-term (lower-frequency) features. As the top panels indicate, the matter memory \adam{final metric displacement} for the SFHo model is larger, reflecting the greater quadrupolar asymmetry and energy of its blast (see Figure \ref{fig:3Dshock}). The magnitude and ``direction" of this for both models is likely random, though its smaller value for the DD2 model likely partially reflects its weaker and more dipolar (as opposed to quadrupolar) explosion.

The bottom panel of Figure \ref{fig:strain} provides an interesting insight into another aspect of the gravitational-wave signatures of CCSNe in general, and the associated differences between these two EOS realizations in particular. It depicts the metric strain \adam{along various directions} due to anisotropic neutrino emissions, the so-called  longer-term/lower-frequency ``neutrino memory" \citep{epstein1978,burrows1996,vartanyan2020,choi2024}. First, we see that this component is generally of greater magnitude than that for the matter (top panels). \adam{This is quite a common facet of CCSN GW signals \citep{choi2024}.} \newadam{ As the referee poi, we note, as expected \citep{choi2024}, that the neutrino memory for these runs is a strong function of direction. The quadrupolar part of the angular dependence of the neutrino emissions does change with time in complicated ways and its principal axes do change direction. However, it is possible that those principal axes can settle down, with the result that over time a particular direction experiences significantly less neutrino memory. This seems to be what we witness in the bottom panel of Figure \ref{fig:strain}. We note, however, that Earth is unlikely to be along that particular direction when the next galactic supernova occurs, so surveying the bottom panel in an average sense is reasonable.} \newadam{Third}, we see that this component is weaker for the DD2 run (dashed) than for the SFHo run (solid), \newadam{though only in the average sense}.  The relatively lower magnitude for the neutrino memory of the DD2 model likely reflects its lower early-time neutrino luminosity (Figure \ref{fig:lum}) and lower degree of matter (and, hence, neutrino emission) quadrupolar asymmetry. While one must acknowledge the imperfections in our neutrino transport approach (though it is multi-D, see \citep{skinner2019}) and the physical reality of expected variations due to the chaotic character of supernova hydrodynamics (and, hence, radiation/hydrodynamics), the relative behavior vis \`a vis the differences we witness in these neutrino memory signatures is suggestive, if not definitive. We point out that this hierarchy in the neutrino memory magnitude with neutrino luminosity is echoed in our more detailed study of neutrino memory \citep{choi2024}, \adam{which also provides the neutrino memory as a function of direction}.

\begin{figure*}   
    \centering
    \includegraphics[width=0.47\textwidth]{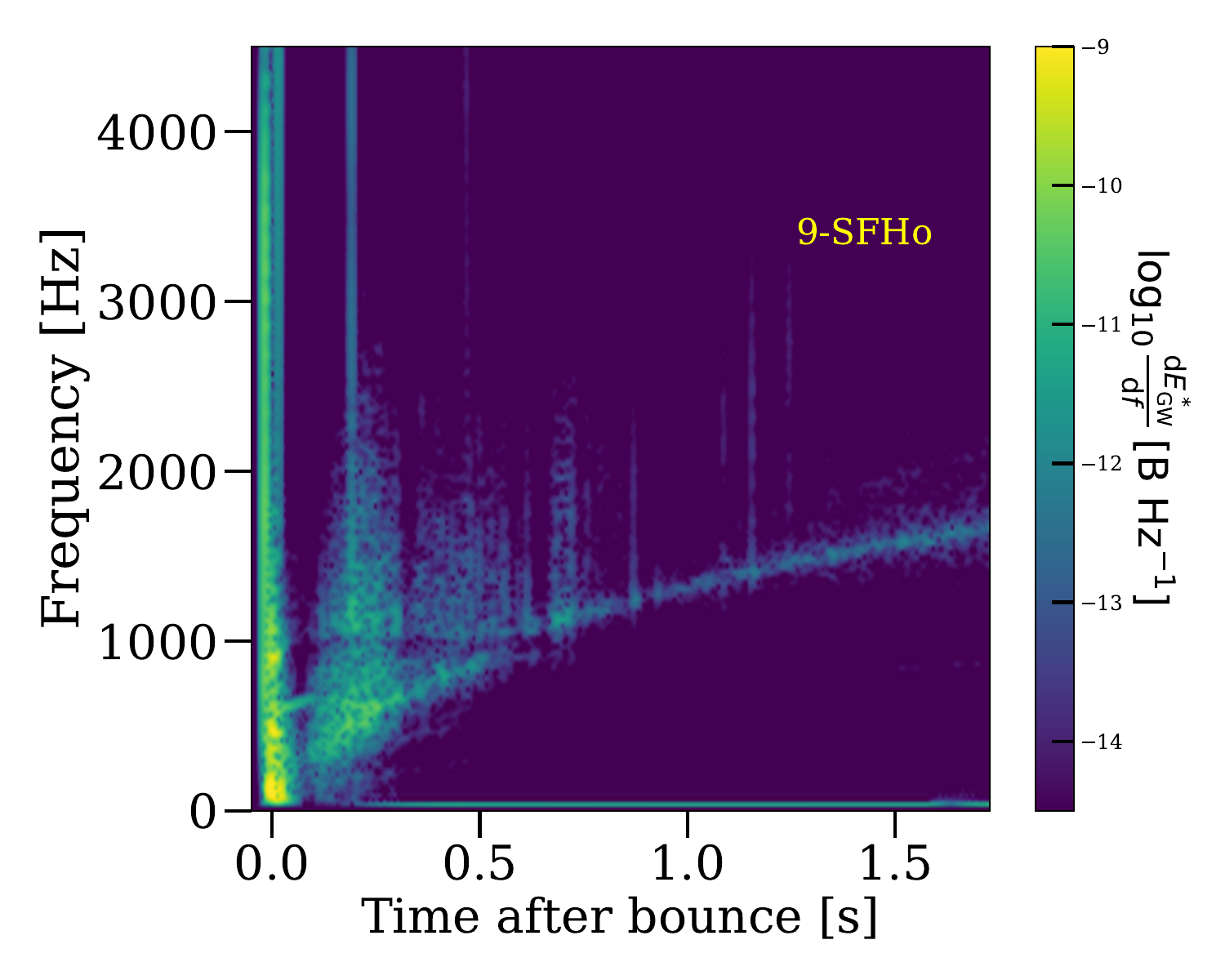}
    \includegraphics[width=0.47\textwidth]{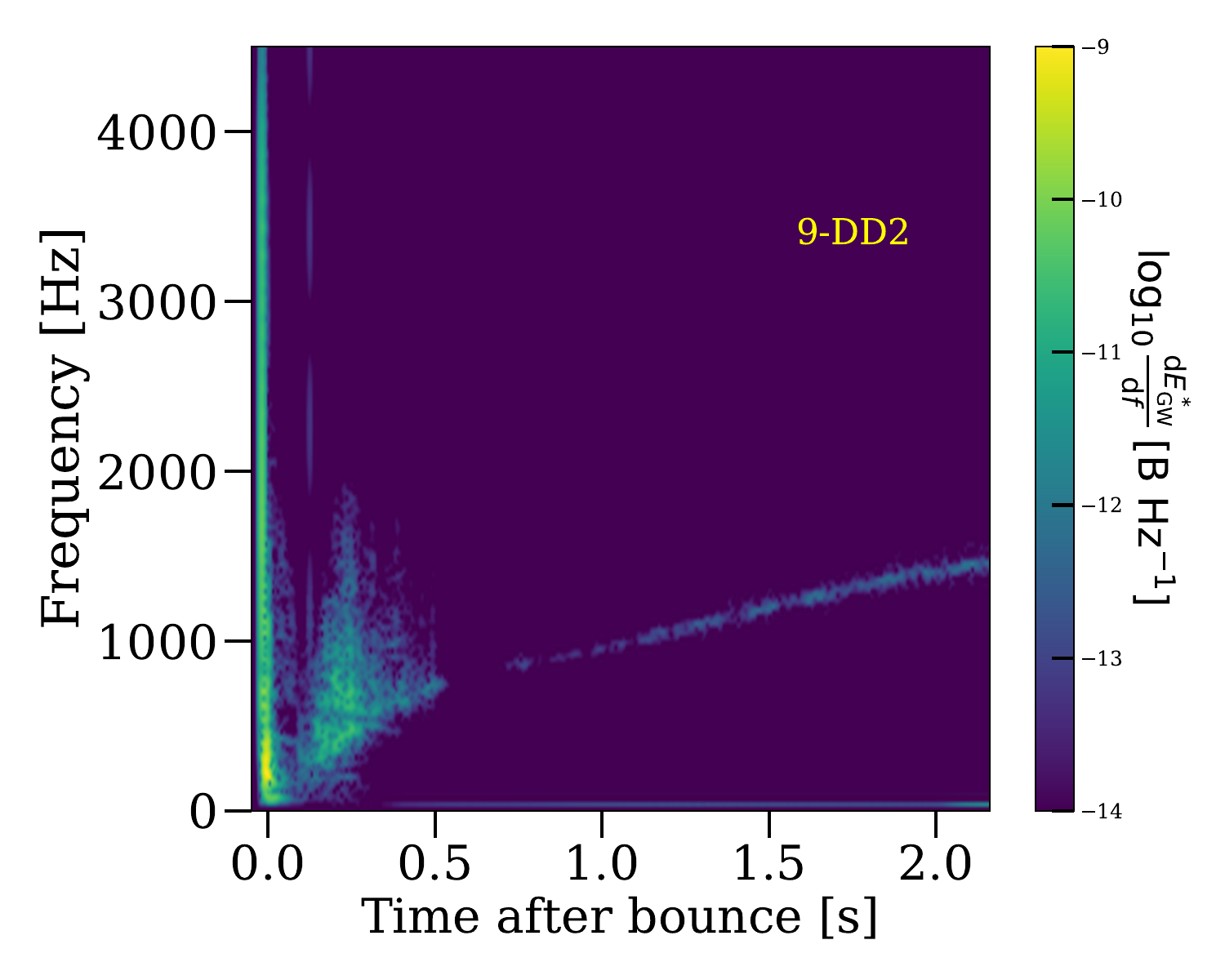}
    \caption{Comparison of the gravitational-wave energy spectrograms for both the SFHo (left) and DD2 (right) equations of state. The SFHo EoS has significantly more power in the later-time f-mode, which also has a higher frequency than found using the DD2 EOS. Note also the lower frequency (near $\sim$600 Hz) for the \newadam{power gap} for the DD2 EOS than seen for the SFHo EOS (near $\sim$1000 Hz). Curiously, for the SFHo EOS, the initial branch of what will later \newadam{merge} into its f-mode starts \newadam{near the power gap} regime of the DD2 simulation. \newadam{Clearly, a full modal analysis is called for.} \neweradam{Note that the matter gravitational-wave data after the last restart of these simulations was corrupted, and hence is not plotted.  However, less than $\sim$1\% of the emission occurs after these times.} See text for a discussion.}
    \label{fig:spectrogram}      
\end{figure*}

We have generated the corresponding spectrograms (frequency-time plots) and display the results in Figure \ref{fig:spectrogram}. The distinctive g/f-mode feature \citep{2018ApJ...861...10M} and its time evolution are clearly evident, as is the higher frequency ``haze," also seen in \citet{vartanyan2023}. We note that \neweradam{a high sampling rate is needed to capture the ``haze" component that trends to high frequencies.} This haze is due to the rapid and stochastic deceleration at the PNS of the numerous infalling matter plumes and can constitute many tens of percent of the total radiated GW energy. \newadam{For our models here this component is roughly 10\%, but for more massive and higher-compactness progenitors can be more than 50\% \citep{vartanyan2023}.} The rapid deceleration is almost like a delta function in real time, that when Fourier-analyzed spreads into a broad-spectrum signal in frequency space. Our sampling cadence is \newadam{every forty timesteps, which translates into a rate of} $\sim$25 kHz to $\sim$50 kHz (Nyquist-limited at half this frequency). \newadam{We note that though our 9-$M_{\odot}$ model calculations reveal power to $\sim$3 KHz, our other models at higher progenitor ZAMS mass and compactness evince power all the way up to $\sim$8 KHz \citep{vartanyan2023}.}

Importantly, \newadam{we find that} the f-mode frequency is slightly higher for the SFHo model, reflecting the more compact nature of its PNS core \citep{2018ApJ...861...10M}. We also see the very early-time (first $\sim$50 milliseconds) ``prompt" convection signature that spans a very broad range of frequencies.  This component is stronger than seen in \citet{vartanyan2023} due to the fact that the collapse phase of the calculations here was done in 3D, allowing numerical seed perturbations to grow to larger amplitudes at given post-bounce times than when this phase is simulated in 1D, and then mapped to 3D 10 milliseconds after bounce, as was the case in \citet{vartanyan2023}.  

However, one of the most curious aspects of our 3D EOS comparison study is the position and character of the regions of the spectrograms associated with mixed modes. As \adam{suggested} in both \citet{2018ApJ...861...10M} and \citet{vartanyan2023}, there \newadam{seem to be} avoided crossings, \neweradam{here near $\sim$0.5-0.6 seconds after bounce}, due to the repulsion of a trapped core g-mode and the f-mode with the same ``quantum numbers" (see also \citet{jakobus}). However, the \newadam{nearly horizontal ``power gaps" \citep{vsg2018,eggenberger2021,vartanyan2023,andresen2026}} are at very different frequencies for the two different equations of state (at $\sim$550 Hz for the DD2 and $\sim$1000 Hz for the SFHo). \neweradam{By power gap \citep{andresen2026} we mean the nearly horizontal features in Figure \ref{fig:spectrogram} where there is a dearth of power. The avoided crossings and the power gaps seem to be related, since the power gap ends where the avoided crossing occurs, but the reason for this is unclear and further work is clearly necessary to explain this behavior.} Furthermore, \newadam{where we see the gap for the DD2 EOS we see a filled horizontal signal for the SFHo EOS.} \neweradam{The reason for this behavior is an open question. In part, it likely reflects the different modal structures and convection profiles of the PNS for the different equations of state and may provide an interesting EOS diagnostic. However, a comprehensive modal analysis is clearly called for.} \newadam{We note that \citet{eggenberger2021},} using their 2D simulations, found that the frequency of the power gap is mostly a reflection of the effective mass of the nucleon ($m^*$) and that higher $m^*$s lead to higher frequencies. They also note a dependence on the central density in both the f-mode \newadam{and power gap}, something we also see. We stress that important determinants of the gravitational-wave spectrogram are indeed the eigenvalues and eigenfunctions of the excitable modes. However, these are affected not just directly by the nuclear EOS, but indirectly by which spatial regions are convective and, hence, cannot support g-modes. Hence, the differences in the behavior of PNS convection we described in \S\ref{hydro} and depict in Figure \ref{fig:conv_flux} suggest caution in the interpretation of these modes as a one-dimensional diagnostic. Therefore, while the mixed-modal gravitational-wave features in CCSNe are fascinating potential diagnostics of the nuclear equation of state, one must have a holistic (if more complicated) perspective on their diagnostic possibilities, that includes the behavior and extent of PNS convection. It will be important to more fully explore this intriguing topic in the future, \adam{in particular, \newadam{as noted above,} with a comprehensive modal analysis \citep{afle2023}} in the context of the broad spectrum of core-collapse supernova progenitor stars.  

\adam{Finally, we note that, given the equation of state dependence of the GW signatures, one should be cautious in mixing signatures with different EOSes when extracting trends along the progenitor and supernova property continuum. This was carefully done in \citet{radice2019} and \citet{vartanyan2023}, but can result in erroneous scaling laws if models with different EOSes are combined in such an exercise \citep{lella2026}.}

\section{Discussion and Conclusions}
\label{conclusions}

With this paper, and using a state-of-the-art simulation code, we have embarked upon the study of the dependence on the nuclear equation of state of the asymptotic state observables of core-collapse supernova explosions in three spatial dimensions.  In the past, studies have focused on explodability \citep[e.g.,][]{Couch2013,Schneider2019,PhysRevLett.124.092701,Ghosh2022,Powell2025}, but have not been carried out both in 3D and to late enough times to witness the final explosion energies, nucleosynthesis, recoil kicks, and gravitational wave signatures. As anticipated, an EOS which results in a more extended protoneutron star core explodes, if it does, with a lower energy and diminished $^{56}$Ni yield, as well as a likely lower kick speed.  We do not see a clean discrimination of the stiffness \citep{Couch2013} and $m^*$ \citep{Schneider2019,PhysRevLett.124.092701} effects, though both likely play a role. In addition, and as expected, the neutrino luminosities are lower for the DD2 model during most of the early explosion phase, as are the mean neutrino energies during much of the post-bounce evolution. Since for a stiffer EOS (such as the DD2) and a given residual baryon mass the total binding energy of the cold, catalyzed neutron star will be lower, we note that the total aggregate neutrino energy lost will be lower as well. Interestingly, we observe in the DD2 model an earlier onset of global PNS convection and an associated temporary enhancement (for the 9-$M_{\odot}$ model near one second after bounce) in the electron neutrino luminosity due to an enhanced episode of convective lepton and energy transport from the inner core. A dependence upon the temporal and quantitative behavior of PNS convection is reflected not just in the neutrino emissions, but in
the characteristics of the gravitational wave signature. We have seen when comparing the SFHo and DD2 models a distinct difference in the frequencies of \newadam{the avoided crossings, the later-time f-mode frequency ($\sim$15\%), and the ``power gap''}, first seen in \citet{vsg2018}).

The differences in the nucleosynthesis reflect the relative speed with which the explosion is ignited and with which the shock is launched, with the faster exploding model (SFHo) yielding more species with atomic weights between 60 and 90. This has a 
bearing on the environments in which a weak r-process might occur in CCSN explosions. However, as demonstrated in \citet{wang_nucleo_2024}, this is perhaps more directly connected to the presence and magnitude of seed perturbations in turbulent lower-mass progenitors and is not so cleanly connected to the nuclear EOS. Though we think it reasonable to conclude that faster exploding models for which the $Y_e$ can freeze-out with a larger neutron-rich component are more likely sites of the weak r-process (most likely for the lower-mass massive star CCSN progenitors), we have not here demonstrated a clean connection between weak r-process yields and the EOS.

\begin{figure*}   
    \centering
    \includegraphics[width=0.65\textwidth]{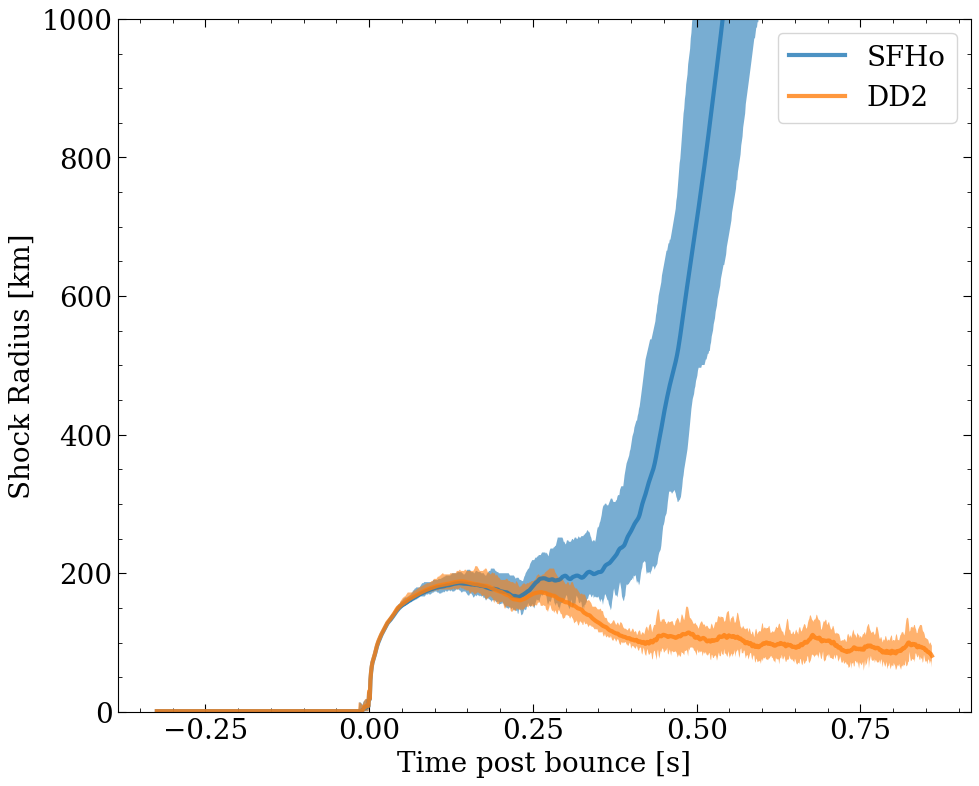}
    \caption{Mean (solid line), maximum, and minimum shock radii for the 18.5 $M_{\odot}$ progenitor star versus time after bounce (in seconds) for the SFHo (blue) and DD2 (orange) equations of state on a scale of 1000 km. Here, there is a qualitative difference in the outcome, where the SFHo model explodes, but the DD2 model does not. For the DD2 EOS, the correspondingly lower rate of heat deposition in the gain region for this higher-compactness 18.5-$M_{\odot}$ progenitor model is not adequate to overcome the more rapid infall accretion ram pressure experienced at its stalled bounce shock \citep{wang}.}
    \label{fig:shock_r_comp.18.5}      
\end{figure*}

We emphasize that this paper is merely a first attempt to explore the explosion observables in the context of sophisticated late-time 3D CCSN models. What is here most glaringly missing is a comprehensive study in 3D to late times of the EOS dependence as a function of progenitor mass. How do the explodability and diagnostic observables vary for the range of viable equations of state and progenitor masses
from 9 to $\sim$80 solar masses and a range of metallicities. Figure \ref{fig:shock_r_comp.18.5} shows our preliminary 3D SFHo/DD2 EOS exploration of the outcome of collapse for an 18.5-$M_{\odot}$ progenitor.  We see that, though the SFHo model explodes, the DD2 model does not (!). This is a qualitative difference, that will also bear on the range of progenitors that leave behind black holes instead of neutron stars. For the DD2 EOS, the neutrino heating rate in the gain region is likely lower for all progenitors. However, does this merely delay a weaker explosion for many progenitors, and leave behind a neutron star (of perhaps higher mass)? Do more progenitors not explode? Or do more DD2 models explode, but leave behind black holes, such as we saw for the 23-$M_{\odot}$ model of \citet{sukhbold2018} using our supernova code F{\sc{ornax}} \citep{Channels2025}? And, more to the point, how does the residual black hole mass spectrum depend on the nuclear EOS? It is true that the DD2 EOS boasts a higher maximum neutron star mass and generically larger neutron star radii than are currently suggested by data \citep{Nicer2019,Rutherford2024} and that the SFHo EOS model is likely closer to reality \footnote{Note that the softer LS220 EOS \citep{1991NuPhA.535..331L} and the stiffer Shen EOS \citep{Shen1998} have already be eliminated.}. Nevertheless, what we see in the set of calculations we provide in this paper demands a closer look into the mapping of the nuclear EOS with supernova, neutron star, and black hole observables than has been conducted to date. This investigation will be expensive, but essential, as the theory community continues to converge on a fully predictive theory of CCSN observables. 


\section*{Data Availability}  

The data presented in this paper can be made available upon reasonable request to the authors.

\section*{Acknowledgments}

We thanks Evan O'Connor for help with the format of the DD2 equation of state tables. TW acknowledges support by the U.~S.\ Department of Energy under grant DE-SC0004658, support by the Gordon and Betty Moore Foundation through Grant GBMF5076, and support through a Simons Foundation grant (622817DK). DV acknowledges support from the NASA Hubble Fellowship Program grant HST-HF2-51520. AB acknowledges former support from the U.~S.\ Department of Energy Office of Science and the Office of Advanced Scientific Computing Research via the Scientific Discovery through Advanced Computing (SciDAC4) program and Grant DE-SC0018297 (subaward 00009650) and former support from the U.~S.\ National Science Foundation (NSF) under Grant AST-1714267. We are happy to acknowledge access to the Frontera cluster (under awards AST20020 and AST21003). This research is part of the Frontera computing project at the Texas Advanced Computing Center \citep{Stanzione2020}. Frontera is made possible by NSF award OAC-1818253. Additionally, a generous award of computer time was provided by the INCITE program, enabling this research to use resources of the Argonne Leadership Computing Facility, a DOE Office of Science User Facility supported under Contract DE-AC02-06CH11357. Finally, the authors acknowledge computational resources provided by the high-performance computer center at Princeton University, which is jointly supported by the Princeton Institute for Computational Science and Engineering (PICSciE) and the Princeton University Office of Information Technology, and our continuing allocation at the National Energy Research Scientific Computing Center (NERSC), which is supported by the Office of Science of the U.~S.\ Department of Energy under contract DE-AC03-76SF00098.

\newpage

\bibliographystyle{aasjournal}
\bibliography{References}


\begin{deluxetable*}{ccc}
\tablecolumns{3}
\tablewidth{0pt}

\begin{minipage}{\textwidth}
  \centering
  \textbf{Basic Model Results} \\  
\end{minipage}

\tablehead{\colhead{Quantity} & \colhead{DD2}& \colhead{SFHo}}
\startdata
Explosion energy [B]       & $0.080$ & $0.123$  \\
$M_{\rm bar} [M_\odot]$ & 1.360 & 1.353\\
$M_{\rm grav} [M_\odot]$ & 1.259 & 1.242\\
$^{56}$Ni [10$^{-3}$$M_{\odot}$] &6.30 &7.26 \\
$^{44}$Ti [10$^{-6}$$M_{\odot}$] &3.59 &4.22 \\
$|\bf{V}|_{\rm kick}$ [km s$^{-1}$]& 68 & 132 \\   
\enddata
\caption{Main asymptotic physical observables of the DD2 and SFHo simulations for the 9-$M_{\odot}$ progenitor from \citet{swbj16}.}
\label{tab:main}      
\end{deluxetable*}

\label{lastpage}

\end{document}